\documentclass[a4paper,11pt]{article}
\pdfoutput=1 

\usepackage{jheppub} 
  
  \newbox\pippobox
  
  \def\({\left(} \def\){\right)}
  \def\[{\left[} \def\]{\right]}

  \def\gz{\mathrel{\mathop g^{\scriptscriptstyle{(0)}}}{}\!}
  
  \def\gs{\mathrel{\mathop g^{\scriptscriptstyle{(2)}}}{}\!}

  \newcommand{\be}{\begin{equation}}
  	\newcommand{\ee}{\end{equation}}
  \newcommand{\bea}{\begin{eqnarray}}
  	\newcommand{\eea}{\end{eqnarray}}
  \newcommand{\ba}{\begin{eqnarray}}
  	\newcommand{\ea}{\end{eqnarray}}

  \newcommand{\beq}{\begin{equation}}
  	\newcommand{\eeq}{\end{equation}}
  \newcommand{\beqa}{\begin{eqnarray}}
  	\newcommand{\eeqa}{\end{eqnarray}}
  \newcommand{\beqar}{\begin{eqnarray*}}
  	\newcommand{\eeqar}{\end{eqnarray*}}

  \newcommand{\mt}[1]{\textrm{\tiny #1}}

  \newcommand{\veps}{\varepsilon}

  \def\de{\delta}

\usepackage[T1]{fontenc} 

\title{\boldmath Universal Terms of Entanglement Entropy for 6d CFTs}

\author[a]{Rong-Xin Miao}

\affiliation[a]{ Max Planck Institute for Gravitational Physics
	(Albert Einstein Institute)\\ Am M\"uhlenberg 1, 14476 Golm,
	Germany}

\emailAdd{rong-xin.miao@aei.mpg.de}

\abstract{We derive the universal terms of entanglement entropy for 6d CFTs by applying the holographic and the field theoretical approaches, respectively. Our formulas are conformally invariant and agree with the results of \cite{Hung,Safdi}. Remarkably, we find that the holographic and the field theoretical results match exactly for the $C^2$ and $Ck^2$ terms, where $C$ and $k$ denote the Weyl tensor and the extrinsic curvature, respectively. As for the $k^4$ terms, we meet the splitting problem of the conical metrics. The splitting problem in the bulk can be fixed by equations of motion. As for the splitting on the boundary, we assume the general forms and find that there indeed exists suitable splitting which can make the holographic and the field theoretical $k^4$ terms match. Since we have much more equations than the free parameters, the match for $k^4$ terms is non-trivial.}

\begin{document} 
\maketitle
\flushbottom

\section{Introduction}

Entanglement entropy (EE) plays an important rule in the fields of gravity \cite{Nishioka} and quantum many-body physics \cite{Amico,Eisert}. It is non-local and provides a useful tool to probe the quantum correlations. EE can be calculated by applying the holographic method \cite{Ryu1,Ryu2} and the perturbative approach \cite{Smolkin}. The holographic entanglement entropy is a rapidly developing field. Recently, the Ryu-Takayanagi conjecture \cite{Ryu1,Ryu2} has been proved by Lewkowycz and Maldacena \cite{Maldacena1}. See also \cite{Fursaev,Casini12} for some early tries. Later, the approach of \cite{Maldacena1} is generalized to higher curvature gravity \cite{Dong,JCamps} and the most general higher derivative gravity \cite{Miao2,Miao3}. See also \cite{Solodukhin1,Erdmenger,1Fursaev1,Bhattacharyya1,Bhattacharyya2,Bhattacharyya3} for the study of holographic entanglement entropy and the differential entropy of a holographic hole \cite{Balasubramanian,Balasubramanian1,1Myers,Headrick}. Other interesting developments include the correspondence between bulk locality and quantum error correction \cite{Almheiri,Polchinski}, the quantum Bousso bound \cite{Bousso:2014sda,Bousso:2014uxa}, the RG flow of entanglement entropy \cite{Rosenhaus:2014nha,Cremonini}, quantum entanglement of local operators \cite{Nozaki:2014hna}, the relation between quantum dimension and entanglement entropy \cite{He:2014mwa} and the holographic three point functions of stress tensor \cite{KallolSen}.

In this paper, we focus on the universal terms of EE. As we know, 
the leading term of EE obeys the area law. However, in spacetime dimensions higher than two, it depends on the cutoff of the system. In contrast to the leading term, the logarithmic term of EE in even spacetime dimensions is universal and thus is of great interest. The logarithmic term of EE for 2d CFTs is given by \cite{Cardy,Wilczek}
\begin{eqnarray}\label{2dlog}
	S_{\Sigma}|_{\log}=\frac{c}{3}\log\big{(}\frac{L}{\pi \delta}\sin(\frac{\pi l}{L})\big{)}
\end{eqnarray}
where $l$ and $L$ are the length of the subsystem and total system, respectively. $\delta$ denotes the cutoff and $c$ is the central charge of the CFT. 

In 4-dimensional space-time, the logarithmic term of EE is proposed by \cite{Solodukhin}
\begin{eqnarray}\label{4dlog}
	S_{\Sigma}|_{\log}=\log(\ell/\delta)\frac{1}{2\pi} \int_{\Sigma} \big{[} c (C^{ijkl}h_{ik}h_{jk}-tr k^2+\frac{1}{2}(tr k)^2)-a R_{\Sigma}\big{]},
\end{eqnarray}
where $C_{ijkl}$ is the Weyl tensor, $k$ is the extrinsic curvature and $R_{\Sigma}$ is the intrinsic Ricci scalar, $a$ and $c$ are the central charges of 4d CFTs. 
Eq.(\ref{4dlog}) is firstly derived by using the holographic entanglement entropy (HEE) of Einstein gravity \cite{Solodukhin}. Later, by applying Dong's formula \cite{Dong}, \cite{Miao1} prove that the general higher curvature gravity yields the same results.   

So far, not much is known about the logarithmic term of EE for 6d CFTs except \cite{Hung, Safdi}. In \cite{Hung}, Hung, Myers and Smolkin (HMS) obtain the logarithmic term of EE for 6d CFTs in case of zero extrinsic curvatures. Because the condition $k_{aij}=0$ breaks the conformal invariance, their formulas are not conformally invariant. In \cite{Safdi}, Safdi studies the cases with $B_3=\frac{B_2-\frac{B_1}{2}}{3}$ in flat space, where $B_i$ are the central charges of 6d CFTs. Since the `flat-space condition' is imposed, the results of \cite{Safdi} are not conformally invariant either. Now let us briefly review their works.

HMS derive the universal terms of EE for CFTs as the `entropy' of its Weyl anomaly \cite{Hung,Myers}. Take the Weyl anomaly as a gravitational action and then calculate the `entropy' of this `action'. It turns out that this 'entropy'  equals to the logarithmic term of EE for CFTs. In six dimensions, the trace anomaly takes the following form
\be
\langle\,T^i{}_i\,\rangle=\sum_{n=1}^3 B_n\, I_n + 2 A \, E_6,
\label{trace6}
\ee
where $E_6$ is the Euler density and $I_i$ are conformal invariants defined by
\bea
I_1&=&C_{k i j l}C^{i m n j}C_{m~~\,n}^{~~k l} ~, \qquad
I_2=C_{i j}^{~~k l}C_{k l}^{~~m n}C^{~~~i j}_{m n}~,\nonumber \\
I_3&=&C_{i k l m}(\nabla^2 \, \de^{i}_{j}+4R^{i}{}_{j}-{6 \over 5}\,R
\,
\de^{i}_{j})C^{j k l m}\,.\label{trace6x}
\eea
For entangling surfaces with the rotational symmetry, only Wald entropy contributes to the holographic entanglement entropy. Thus, we have
\beq
S_\mt{EE} = \log(\ell/\delta)\, \int d^{4}x\sqrt{h}\,\left[\,2\pi\,
\sum_{n=1}^3 B_n\,\frac{\partial I_n}{\partial R^{i j}{}_{k
		l}}\,\tilde{\veps}^{i j}\,\tilde{\veps}_{k l} +2\,A\,
E_{4}\,\right]_{\Sigma}\,,
\label{Waldformula8z}
\eeq
where
\bea
\frac{\partial I_1}{\partial R^{i j}{}_{k l}}\,\tilde{\veps}^{i
	j}\,\tilde{\veps}_{k l} &=& 3 \left(C^{j m n k}\, C_{m~~n}^{~~i
	l}\,\tilde\veps_{i j}\,\tilde\veps_{k l} - \frac{1}{4}\,
C^{iklm}\,C^{j}_{~kl m }\,\tilde{g}^{\perp}_{i j} + \frac{1}{20}\,
C^{ijkl}\,C_{ijkl}\right)
\,,\label{CFTFW1}\\
\frac{\partial I_2}{\partial R^{i j}{}_{k l}}\,\tilde{\veps}^{i
	j}\,\tilde{\veps}_{k l} &=&3\left(C^{k l m n}\, C_{m n}^{~~~ i j}\,
\tilde\veps_{i j}\,\tilde\veps_{k l} -C^{iklm}\,C^{j}_{~kl m
}\,\tilde{g}^{\perp}_{i j} + \frac{1}{5}\, C^{ijkl}\,C_{ijkl}\right)
\,,\label{CFTFW2} \\
\frac{\partial I_3}{\partial R^{i j}{}_{k l}}\,\tilde{\veps}^{i
	j}\,\tilde{\veps}_{k l} &=&  2\left(\Box\, C^{i j k l}+ 4\,R^{i}{}_{m}
C^{m j k l}-\frac{6}{5}\, R\, C^{i j k l}\right) \tilde\veps_{i
	j}\,\tilde\veps_{k l}- 4\, C^{i j k l}\,R_{ik}\,\tilde{g}^{\perp}_{jl}
\nonumber \\
&& \qquad+ 4\,C^{iklm}\,C^{j}_{~kl m }\,\tilde{g}^{\perp}_{i
	j}-\frac{12}{5}\, C^{ijkl}\,C_{ijkl} \, .\label{CFTFW3} 
\eea
For entangling surfaces without the rotational symmetry, the anomaly-like entropy from $C_{ijkl}\Box C^{ijkl}$ should be added to the entropy eq.(\ref{Waldformula8z}). This contribution is used in \cite{Miao2} to resolve the HMS puzzle \cite{Hung}. See \cite{Astaneh1,Astaneh2} for an alternative try, which suggests to use the entropy from total derivative terms to explain the HMS mismatch. However, it is found that the entropy of covariant total derivative terms vanishes \cite{Dong1} by applying the Lewkowycz-Maldacena regularization \cite{Maldacena1}. What is worse, \cite{Miao3} proves that the proposal of \cite{Astaneh1,Astaneh2} fails in solving the HMS puzzle \cite{Miao3} even if the entropy of total derivative terms was non-zero.

Now let us turn to review the work of \cite{Safdi}, where the universal term of EE for 6d CFTs with $B_3=\frac{B_2-\frac{B_1}{2}}{3}$ in flat space is obtained 
\begin{eqnarray}\label{6dlog}
	S_{\Sigma}|_{\log}=\log(\ell/\delta) \int_{\Sigma} 2A E_4+6\pi \big{[}3(B_2-\frac{B_1}{4})J+B_3 T_3\big{]}.
\end{eqnarray}
Here $J=T_1-2T_2$ and $T_i$ are given by
\begin{eqnarray}\label{Ti}
	T_1=(\text{tr} \bar{k}^2)^2,\ T_2=\text{tr} \bar{k}^4, \ T_3=(\nabla_a k)^2-\frac{25}{16}k^4+11k^2 tr k^2-6(tr k^2)^2-16 k tr k^3+12 tr k^4,\nonumber\\
\end{eqnarray}
where $\bar{k}$ denotes the traceless part of the extrinsic curvature. For simplicity, \cite{Safdi} focus on flat space and set the extrinsic curvature in the time-like direction to be zero. Since the `flat-space condition' breaks the conformal invariance, the results of \cite{Safdi} $(T_3)$ are not conformally invariant either.

In this paper, we investigate the most general cases. By applying the holographic and the field theoretical methods respectively, we derive the universal terms of EE for 6d CFTs. Our formulas are conformally invariant and reduce to those of \cite{Hung, Safdi} when imposing the conditions they use. Remarkably, we find that the holographic and the field theoretical results match exactly for the $C^2$ and $Ck^2$ terms, where $C$ and $k$ denote the Weyl tensor and the extrinsic curvature, respectively. As for the $k^4$ terms, we have to deal with the splitting problem of the conical metrics. The splitting problem appears because one can not distinguish $r^2$ and $r^{2n}$ ($n\to 1$) in the expansions of the conical metrics. We can fix the splitting problem in the bulk by applying equations of motion. As for the splitting problem on the boundary, we assume the general expressions and find that there does exist suitable splittings which can make the holographic and the field theoretical $k^4$ terms match. 

It should be mentioned that the splitting problem does not affect the logarithmic term of EE for 4d CFTs. That is because only the $O(K^0)$ and $O(K^2)$ terms ($K$ denote the extrinsic curvature in the bulk) of the entropy contribute to the logarithmic term of EE for 4d CFTs \cite{Miao1}, however these terms are irrelevant to the splitting problem \cite{Miao2}. As for the 6d logarithmic terms, we need to calculate the $O(K^4)$ terms of the entropy, which come from cubic curvature terms in the action. It turns out that the only cubic curvature term irrelevant to the splittings is the Lovelock term. However, the central charges of CFTs dual to Lovelock gravity and the curvature-squared gravity are not independent but constrained by $B_3=\frac{B_2-\frac{B_1}{2}}{3}$. Thus, to study the most general case in 6-dimensional space-time, we have to deal with the splitting problem.

An overview of this paper is as follows:  We begin with a brief review of the holographic entanglement entropy and the discussions of the splitting problem for the conical metrics in Sect. 2. In Sect. 3, we take the general higher curvature gravity as an example to illustrate the holographic approach for the derivations of the universal terms of EE. In Sect. 4, we derive the universal terms of EE for 6d CFTs by applying the holographic method.  We firstly derive the results from a smart-constructed action and then prove that the general action produces the same results. In Sect. 5, we use the field theoretical method to calculate the universal terms of EE for 6d CFTs.  We compare the field theoretical results with the holographic ones and get good agreements. We conclude with a brief discussion of our results in Sect. 6.

Notations: $X^{\mu}$, $\hat{G}_{\mu\nu}$ and $\hat{R}_{\mu\nu\rho\sigma}$ are the coordinates, the metric and the curvature in the $(d+1)$-dimensional bulk, respectively. Similarly, $x^{i}$, $g_{ij}$ and $R_{ijkl}$ denote the coordinates, the metric and the curvature on the $d$-dimensional boundary. The entangling surface (extrinsic curvatures) in the bulk and on the boundary are labeled by $m$ and $\Sigma$ ($K$ and $k$), respectively. $h_{\alpha\beta}$ and $\overset{\scriptscriptstyle{(0)}}{h}_{\hat{i}\hat{j}}$ are the induced metrics on $m$ and $\Sigma$, respectively. Notice that $m$ is a $(d-1)$-dimensional manifold while $\Sigma$ is a $(d-2)$-dimensional manifold. 

\section{Holographic entanglement entropy}

\subsection{Holographic entanglement entropy}

In this section, we briefly review the derivations of holographic entanglement entropy (HEE) for the general higher curvature gravity \cite{Dong}. The basic idea is to apply the replica trick and extend it to the bulk. Let us start with the Renyi entropy
\begin{eqnarray}\label{renyi}
	&&S_n=- \frac{1}{n-1}\log tr[\rho^n]=-\frac{1}{n-1}(\log Z_n-n \log Z_1)\\
	&&Z_n=Tr[\hat{\rho}^n],\ \rho=\frac{\hat{\rho}}{Tr[\hat{\rho}]},
\end{eqnarray}
where $\rho$ is the reduced density matrix associated with the subsystem and $Z_n$ is the partition function of the field theory on a suitable manifold $M_n$ known as the $\text{n-fold}$ cover. For theories with a holographic dual we can build a suitable bulk solution $B_n$ whose boundary is $M_n$. Then the gauge-gravity duality identifies the field theory partition function on $M_n$ with the on-shell bulk action on $B_n$
\begin{eqnarray}\label{Zn}
	Z_n=Z[M_n]=e^{-I[B_n]}.
\end{eqnarray}
We can derive the HEE by taking the limit $n\to 1$ of Renyi entropy
\begin{eqnarray}\label{SHEE}
	S_{\text{HEE}}&=&\lim_{n\to 1}S_{n}=-\partial_n(\log Tr[\rho^n])|_{n\to 1}=-Tr[\rho\log \rho]\nonumber\\
	&=&-\partial_n(\log Z_n-n \log Z_1)|_{n\to 1}=\partial_n(I[B_n]-nI[B_1])|_{n\to 1}\nonumber\\
	&=&-\partial_{\epsilon} I_{reg}|_{\epsilon\to 0},
\end{eqnarray}
where $I_{reg}=(nI[B_1]-I[B_n])$ is the regularized action and $\epsilon=1-\frac{1}{n}$.
To derive $I_{reg}$, one need to regularize the conical metric appropriately. 

The regularized conical metric in a coordinate system adapted to a neighborhood of the conical singularity is given by \cite{Dong}:
\begin{eqnarray}\label{conemet}
	ds^2 = e^{2A} \[dz d\bar{z} + e^{2A} T (\bar{z} dz-z d\bar{z})^2 \] + \(h_{ij} + 2K_{aij} x^a + Q_{abij} x^a x^b\) dy^i dy^j \nonumber\\
	+ 2i e^{2A} U_i \(\bar{z} dz-z d\bar{z}\) dy^i + \cdots \,.
\end{eqnarray}
Here $x^a \in \{z,\bar{z}\}$ denotes orthogonal directions to the conical singularity, and $y^i$ denotes parallel directions.  The regularized warp factor is
\begin{eqnarray}
	A = -\frac{\epsilon}{2} \log (z\bar{z}+b^2)\,,
\end{eqnarray}
Using the replica trick, one can derive the entropy as eq.(\ref{SHEE})
\begin{eqnarray}\label{HEE}
	S_{\text{HEE}}=-\partial_{\epsilon} I_{reg}|_{\epsilon\to 0}
\end{eqnarray} 
where $I_{reg}$ is the gravitational action got from the regularized metric (\ref{conemet}). There are two kinds of terms relevant to the entropy. The first kind is 
\begin{eqnarray}\label{dA}
	&&\hat{R}_{z\bar{z}z\bar{z}}=e^{2A}\partial_z\partial_{\bar{z}}A+...\nonumber\\
	&&\int dz d\bar{z} \partial_z\partial_{\bar{z}}A=-\pi \epsilon.
\end{eqnarray}
It contributes to Wald entropy. The second kind is 
\begin{eqnarray}\label{dAdA}
	&&\hat{R}_{zizj}=2K_{zij}\partial_zA+..., \ \hat{R}_{\bar{z}k\bar{z}l}=2K_{\bar{z}kl}\partial_{\bar{z}}A+...\nonumber\\
	&&\ \ \ \ \ \ \ \ \ \ \ \ \ \ \int dz d\bar{z} \partial_zA\partial_{\bar{z}}Ae^{-\beta A}=-\frac{\pi \epsilon}{\beta} .
\end{eqnarray}
This is the would-be logarithmic term and contributes to the anomaly-like entropy \cite{Dong}. 

Applying eqs.(\ref{HEE},\ref{dA},\ref{dAdA}), one can derive HEE for general higher curvature gravity $f(\hat{R}_{\mu\nu\sigma\rho})$ \cite{Dong}
\begin{eqnarray}\label{HEEDong}
	S_{\text{HEE}}&=&\frac{1}{8 G}\int d^{d-1}y \sqrt{h}\big{[}\ \frac{\partial f}{\partial \hat{R}_{z\bar{z}z\bar{z}}}
	+16\sum_{\beta}(\frac{\partial^2 f}{\partial \hat{R}_{zizj}\partial \hat{R}_{\bar{z}k\bar{z}l}})_{\beta}\frac{K_{zij}K_{\bar{z}kl}}{\beta+2} \ \big{]}
\end{eqnarray}
Here $\beta$ come from the formula (\ref{dAdA}), and $(\frac{\partial^2 f}{\partial \hat{R}_{zizj}\partial \hat{R}_{\bar{z}k\bar{z}l}})_{\beta}$ are the coefficients in the expansions
\begin{eqnarray}\label{HEEbeta}
	\frac{\partial^2 f}{\partial \hat{R}_{zizj}\partial \hat{R}_{\bar{z}k\bar{z}l}}=\sum_{\beta}e^{-\beta A}(\frac{\partial^2 f}{\partial \hat{R}_{zizj}\partial \hat{R}_{\bar{z}k\bar{z}l}})_{\beta}.
\end{eqnarray}

\cite{Dong} proposes to regularize $Q_{z\bar{z}ij}$ as $e^{2A}Q_{z\bar{z}ij}$. However, as we find in Sect. 4.2, this ansatz yields inconsistent results for the universal terms of EE for 6d CFTs \cite{Miao2}, i.e., two gravitational actions with the same holographic Weyl anomaly give different universal terms of EE. To resolve this inconsistency, we proposes the following regularizations
\begin{eqnarray}\label{ReviewTVQ}
	&&T=e^{-2A}T_0+T_{1},\nonumber\\
	&&Q_{\ z \bar{z} ij}=Q_{0\ \ z \bar{z}ij}+e^{2A}Q_{1\ z\bar{z}ij}
\end{eqnarray}
How to split $M$ into $M_0$ and $M_1$ ($M$ denotes $T$ and $Q$) is the so-called the splitting problem. It appears because one cannot distinguish $r^2$ and $r^{2n}$ in the expansions of the conical metric. It is expected that the splitting problem can be fixed by using equations of motion. As we shall show in the next sub-section, this is indeed the case at least for Einstein gravity.

\subsection{The splitting problem}
Let us investigate the splitting problem in this section. As we have mentioned in the above sub-section, the splittings of the conical metrics cannot be avoided in order to derive consistent results for the universal terms of EE. Actually, the splitting problems appear naturally since we can not distinguish $r^2$ and $r^{2n}$ in the expansions of conical metrics. That is because $r^2$ and $r^{2n}$ become the same order in the limit $n\to 1$ when we calculate HEE. According to \cite{Dong, Camps}, the general regularized conical metric takes the form
\begin{eqnarray}\label{cone}
	ds^2=e^{2A}[dzd\bar{z}+T(\bar{z}dz-zd\bar{z})^2]+2i V_i(\bar{z}dz-zd\bar{z})dy^i\nonumber\\
	+( h_{ij}+Q_{ij})dy^idy^j,
\end{eqnarray}
where $h_{ij}$ is the metric on the transverse space and is independent of $z, \bar{z}$. $A=-\frac{\epsilon}{2}\lg(z\bar{z}+b^2)$ is regularized warp factor. Inspired by \cite{Dong,Camps}, we propose to split  $T, V_i, Q_{ij}$ as
\begin{eqnarray}\label{TVQ}
	&&T=\sum_{n=0}^{\infty}\sum_{m=0}^{P_{a_1...a_n}+1} e^{2 m A }T_{m\ a_1...a_n}x^{a_1}...x^{a_n},\nonumber\\
	&&V_i=\sum_{n=0}^{\infty} \sum_{m=0}^{P_{a_1...a_n}+1} e^{2 m A }V_{m\ a_1...a_n i}x^{a_1}...x^{a_n},\nonumber\\
	&&Q_{ij}=\sum_{n=1}^{\infty} \sum_{m=0}^{P_{a_1...a_n}} e^{2 m A }Q_{m\ a_1...a_n ij}x^{a_1}...x^{a_n}.
\end{eqnarray}
Here $z,\bar{z}$ are denoted by $x^{a}$ and $P_{a_1...a_n}$ is the number of pairs of
$z,\bar{z}$ appearing in $a_1...a_n$. For example, we have $P_{zz\bar{z}}=P_{z\bar{z}z}=P_{\bar{z}zz}=1$, $P_{z\bar{z}z\bar{z}}=2$ and $P_{zz...z}=0$.
Expanding $T,V,Q$ to the first few terms in the notations of \cite{Dong}, we have
\begin{eqnarray}\label{TVQ1}
	&&T=T_0+e^{2A}T_{1}+O(x),\nonumber\\
	&&V_i=U_{0 \ i}+e^{2A}U_{1\ i}+O(x),\nonumber\\
	&&Q_{ij}=2K_{aij}x^a+Q_{0\ abij}x^ax^b+2e^{2A}Q_{1\ z\bar{z}ij}\ z \bar{z}+O(x^3)
\end{eqnarray}
How to split $W$ ($W$ denote $T, V, Q$) into $\{W_{0}, W_{1}, ..., W_{P+1}\}$ is an important problem. It should be mentioned that the splitting problem is ignored in the initial works of Dong and Camps \cite{Dong,Camps}. However they both change their mind and realize the splitting is necessary later \footnote{We thank Dong and Camps for discussions on this problem.}. Recently Camps etal generalize the conical metrics to the case without $Z_n$ symmetry, where the splitting problem appears naturally \cite{Camps1}. Our metric eq.(\ref{cone}) can be regarded as a special case of  \cite{Camps1} with $Z_n$ symmetry. Inspired by \cite{Maldacena1},  it is expected that the splitting problem can be fixed by equations of motion. Let us take Einstein gravity in vacuum as an example.  We denote the equations of motion by $E_{\mu\nu}=\hat{R}_{\mu\nu}-\frac{\hat{R}-2\Lambda}{2}\hat{G}_{\mu\nu}=0$ . Focus on terms which are important near $x^a=0$, we have
\begin{eqnarray}\label{curvature}
	\hat{R}_{ab}&=&2K_{(a}\nabla_{b)}A -\hat{G}_{ab}K^c\nabla_cA+e^{2A}[(12T_1+4 U^2)\hat{G}_{ab}-Q_{1\ abi}^{\ \ \ \ \ \ i}]\nonumber\\
	&&+K_{aij}K_{b}^{\ ij}+(12T_0+8 U_{0}U_{1}) \hat{G}_{ab}-Q_{0\ abi}^{\ \ \ \ \ \  i}\nonumber\\
	\hat{R}_{ai}&=&3\varepsilon_{ba}V^{b}_{\ i}+D^mK_{ami}-D_i K_a, \nonumber\\
	\hat{R}_{ij}&=&r_{ij}+8U_iU_j-Q_{1\ aij}^{\ a}+e^{-2A}[2K_{aim}K^{am}_{\ \ \ j}-K^aK_{aij}+16 U_{0\ (i}U_{1\ j)} -Q_{0\ aij}^{\ a}],\nonumber\\
	\hat{R}&=&r+16U^2+24T_1-2Q_{1\ a\ i}^{\ a\ i}+e^{-2A}(3K_{aij}K^{aij}-K^aK_a+24T_0-2Q_{0\ a\ i}^{\ a\ i}+32U_0U_1),\nonumber\\
\end{eqnarray}
where $A=-\frac{\epsilon}{2} \log z\bar{z} $, $\varepsilon_{z\bar{z}}=\frac{i}{2}$ and $g_{z\bar{z}}=\frac{1}{2}$. Let us firstly consider the leading term of $E_{zz}$, we get
\begin{eqnarray}\label{Ezz}
	E_{zz}=2K_z \nabla_z A+...=-\epsilon \frac{K_z}{z}+...=0.
\end{eqnarray}
Requiring the above equation to be regular near the cone, we obtain the minimal surface condition $K_z=K_{\bar{z}}=0$ \cite{Maldacena1}. To derive $T_0$ and $Q_0$, we need consider the sub-leading terms of $E_{z\bar{z}}, E_{ij}$ and $E_{\mu}^{\mu}$. We have
\begin{eqnarray}\label{Eij}
	&&E_{z\bar{z}}=e^{2A}(...)+[Q_{0\ z\bar{z}i}^{\ \ \ i}-2K_{zij}K_{\bar{z}}^{\ ij}+K_z K_{\bar{z}}-4U_0 U_1]=0,\nonumber\\
	&&E_{ij}=(...)+e^{-2A} [2K_{aim}K^{am}_{\ \ \ j}-K^aK_{aij}+16 U_{0\ (i}U_{1\ j)} -Q_{0\ aij}^{\ a}\nonumber\\
	&&\ \ \ \ \ \ \ \ -\frac{1}{2}h_{ij}
	(3K_{aij}K^{aij}-K^aK_a+24T_0-2Q_{0\ a\ i}^{\ a\ i}+32U_0U_1) ]=0,\nonumber\\
	&&E_{\ \mu}^{\mu}=(...)+\frac{2-D}{2}e^{-2A}[3K_{aij}K^{aij}-K^aK_a+24T_0-2Q_{0\ a\ i}^{\ a\ i}+32U_0U_1]=0.
\end{eqnarray}
Here $ (...)$ denote the leading terms which can be used to determine $T_1, U_{1 i}, Q_{1 z\bar{z}ij}$ and $h_{ij}$. From the subleading terms of the above equations, we can derive a unique solution
\begin{eqnarray}\label{splittingEinstein}
	&&T_0=\frac{1}{24}(K_{aij}K^{aij}-K_aK^a),\nonumber\\
	&&Q_{0z\bar{z}ij}=(K_{zim}K_{zj}^{\ \ m}-\frac{1}{2}K_z K_{\bar{z}ij}+c.c.)+4 U_{0\ (i}U_{1\ j)}
\end{eqnarray}
As we shall show below, a natural choice would be $U_{0\ i}=0$. It should be mentioned that eq.(\ref{splittingEinstein}) are also solutions to the general higher derivative gravity if we require that the higher derivative gravity has an AdS solution. In the next section, we shall use eq.(\ref{splittingEinstein}) to derive the universal terms of EE for 6d CFTs. Actually, we only need a weaker condition near the boundary 
\begin{eqnarray}\label{splittingEinstein1}
	&&T_0=\frac{1}{24}(K_{aij}K^{aij}-x\  K_aK^a)+O(\rho^2),\\
	&&Q_{0z\bar{z}ij}=K_{zim}K_{zj}^{\ \ m}-y\  K_z K_{\bar{z}ij}-z \ g_{ij}K_zK_{\bar{z}}+c.c+O(\rho)\label{splittingEinstein2}
\end{eqnarray}
with $x, y, z$ are some constants which are not important. Here $\rho$ is defined in the Fefferman-Graham expansion eq.(\ref{FGgauge}) and $\rho\to 0$ corresponds to the boundary. Actually, as we shall show in sect.3.2, eq.(\ref{splittingEinstein1}) is the necessary condition that all the higher derivative gravity in the bulk gives consistent results of the universal terms of EE. 

To end this section, let us make some comments. In addition to the equations of motion, there are several other constraints which may help to fix the splitting. Let us discuss them one by one below.

{\bf{1.}} The entropy reduces to Wald entropy in stationary spacetime.

Let us take $\nabla_{\mu}\hat{R}_{\nu\rho\sigma\alpha}\nabla^{\mu}\hat{R}^{\nu\rho\sigma\alpha}$ as an example. In stationary spacetime, we have $K_{aij}=Q_{zzij}=Q_{\bar{z}\bar{z}ij}=0$. Applying the method of \cite{Miao2}, we can derive the HEE as
\begin{eqnarray}\label{ConstraintWald}
	S_{HEE}=S_{Wald}+\int dy^{D-2}\sqrt{h} 128\pi ( Q_{0z\bar{z}ij}Q_{0z\bar{z}}^{\  \ ij}+9 T_0^2+5 (U_{0\ i}U_0^{\ i})^2+ \text{mixed terms of }T_0, Q_0, U_0 ).\nonumber\\
\end{eqnarray}
To be consistent with Wald entropy, we must have $T_0=U_{0\ i}=Q_{0z\bar{z}ij}=0$ in stationary spacetime. This implies that $T_0, U_{0\ i}$ and $Q_{0z\bar{z}ij}$ should be either zero or functions of the extrinsic curvatures.  This is indeed the case for the splitting eqs.(\ref{splittingEinstein}).  By dimensional analysis, we note that $U_{0\ i}\sim O(K)$. However, it is impossible to express $U_{0\ i}$ in terms of the extrinsic curvature $K_{aij}$. Thus, a natural choice would be $U_{0\ i}=0$.

{\bf{2.}}  The entropy of conformally invariant action is also conformally invariant.

In the bulk, we can use gravitational equations of motion to fix the splittings of conical metrics. However, we do not have dynamic gravitational fields on the boundary. Then how can we determine the splittings on the boundary? For the cases with gravity duals, in principle, we can derive the conical metric on the boundary from the one in the bulk. As for the general cases, we do not know how to fix the splittings. If we focus on the case of CFTs, the conformal symmetry can help. As we know, the universal terms of EE for CFTs are conformally invariant. Recall that we can derive the universal terms of EE as the entropy of the Weyl anomaly \cite{Hung,Solodukhin,Myers}. Thus, the entropy of conformal invariants (Weyl anomaly) must be also conformally invariant. Let us call this condition as the `conformal constraint' . Expanding the Weyl tensor in powers of $e^{2A}$, we have
\begin{eqnarray}\label{ConstraintWeyl}
	C_{z\bar{z}z\bar{z}}&=&e^{4A}C_{1\ z\bar{z}z\bar{z}}+e^{2A}C_{0\ z\bar{z}z\bar{z}}\nonumber\\
	C_{zi\bar{z}j}&=&e^{2A}C_{1\ zi\bar{z}j}+C_{0\ zi\bar{z}j},\nonumber\\
	C_{ikjl}&=&C_{1\ ikjl}+e^{-2A}C_{0\ ikjl}
\end{eqnarray}
The `conformal constraint' requires that both $C_1$ and $C_0$ are conformally invariant. Assuming the general splittings in 6d spacetime
\begin{eqnarray}\label{ConstraintWeylSplitting}
	&&T_0=z_1 K_{a mn}K^{amn}+z_2 K_a K^a\\
	&&Q_{0\ z\bar{z}ij}=(x_1 K_{zim}K^{\ m}_{\bar{z}\ \ j}+x_2\  g_{ij} K_{z mn}K_{\bar{z}}^{\ mn}+y_1 K_z K_{\bar{z}ij}+y_2\  g_{ij}K_zK_{\bar{z}}) +c.c.\label{ConstraintWeylSplitting1}
\end{eqnarray}
By using the `conformal constraint', we get
\begin{eqnarray}\label{ConstraintWeylSplittingsolution}
	x_1=1-2y_1,\  x_2=\frac{1}{4}-6 z_1-\frac{y_1}{3},\ y_2=-\frac{1}{16} -6 z_2-\frac{y_1}{24}.
\end{eqnarray}
Thus the `conformal constraint' cannot fix the splittings on the boundary completely.

{\bf{3.}} The splittings should yield the correct universal terms of EE for CFTs.

Another natural constraint for the splittings on the boundary is that it should give the correct universal term of EE for CFTs. By `correct', we mean it agrees with holographic results. Remarkably, the splitting problem does not affect the universal terms of EE for 4d CFTs . From the viewpoint of CFTs, we can derive the universal terms of EE as the entropy of the Weyl anomaly. In 4d spacetime, the Weyl anomaly are curvature-squared terms whose entropy can not include $T_0$ and $Q_0$ by using Dong's formula \cite{Dong}. From the viewpoint of holography, the situation is similar. For the general higher derivative gravity $S(g, R)$, it has been proved that $T_0$ and $Q_0$ does not contribute to the logarithmic terms of EE \cite{Miao1}. As for the 6d CFTs, the splitting problems do matter. To be consistent with the holographic results, in sect. 4, we shall derive the splittings eq.(\ref{ConstraintWeylSplitting}) with 
\begin{eqnarray}\label{ConstraintCFTSplittingsolution}
	x_1=1,\  x_2=\frac{1}{4}-6 z_1,\ y_1=0, \ y_2=-\frac{1}{16} -6 z_2.
\end{eqnarray}
This constraint is better than the `conformal constraint' but still could not fix the splittings completely. It seems that we have some freedom to split the conical metrics on the boundary and this freedom does not affect the universal terms of EE. 

{\bf{4.}} The splittings does not affect the entropy of Lovelock gravity and topological invariants. 

Lovelock gravity is special in several aspects. In particular, it becomes topological invariant in critical dimensions. Thus the entropy of Lovelock gravity must be also topological invariant in critical dimensions. This strong constrains the possible form of the entropy of Lovelock gravity. We know the answer is the Jacobson-Myers formula \cite{Jacobson}. In general, we would get different entropy from the conical metrics with different splittings. Thus, we must check if the splittings affect the entropy of Lovelock gravity. It is clear that the splittings does not affect the Wald entropy. Thus, we focus on the anomaly-like entropy $K_{zij}K_{\bar{z}kl}\frac{\partial^2 L}{\partial \hat{R}_{zizj}\partial \hat{R}_{\bar{z}k\bar{z}l}}$ \cite{Dong}. Note that $T_0$ and $Q_0$ only appear in the curvatures $\hat{R}_{z\bar{z}z\bar{z}}$ and $\hat{R}_{zi\bar{z}j}$ but not $\hat{R}_{ijkl}$. While only $\hat{R}_{ijkl}$ can appear in $\frac{\partial^2 L}{\partial \hat{R}_{zizj}\partial \hat{R}_{\bar{z}k\bar{z}l}}$ for Lovelock gravity. Thus the splittings indeed do not affect the entropy of Lovelock gravity. 

\section{The universal terms of EE}

\subsection{Approach based on PBH transformation}

In this subsection, we introduce an elegant approach \cite{Theisen2}, which rests on the so-called Penrose-Brown-Henneaux (PBH) transformations, for the derivations of the universal terms of EE for CFTs. Taking advantage of the bulk diffeomorphisms and the reparametrization of the world volume (entropy functional), \cite{Theisen2} finds that one do not need to solve equations of motion and the extremal entropy surface in order to derive the universal terms of EE for 4d CFTs. As we shall show in Sect. 4, this is also the case for 6d CFTs. Notice that making no use of equations of motion does not mean the approach of \cite{Theisen2} is off-shell. Actually,  \cite{Theisen2} indeed use some on-shell conditions, which can be derived by applying either PBH transformations or equations of motion. Below we give a brief review of the general method. 

\subsubsection{Fefferman-Graham expansion}
For asymptotically AdS space-time, we can expand the bulk metric in the Fefferman-Graham gauge
\begin{eqnarray}\label{FGgauge}
	ds^2=\hat{G}_{\mu\nu}dX^{\mu}dX^{\nu}=\frac{1}{4\rho^2}d\rho^2+\frac{1}{\rho}g_{ij}dx^idx^j,
\end{eqnarray}
where $g_{ij}=\overset{\scriptscriptstyle{(0)}}{g}_{ij}+\rho \overset{\scriptscriptstyle{(1)}}{g}_{ij}+...+\rho^{\frac{d}{2}}(
\overset{\scriptscriptstyle{(\frac{d}{2})}}{g}_{ij}+\overset{\scriptscriptstyle{(\frac{d}{2})}}{\gamma}_{ij}\log \rho)+...$. Interestingly,
\begin{eqnarray}\label{g1}
	\overset{\scriptscriptstyle{(1)}}{g}_{ij}=-\frac{1}{d-2}(\overset{\scriptscriptstyle{(0)}}{R}_{ij}-\frac{\overset{\scriptscriptstyle{(0)}}{R}}{2(d-1)}\overset{\scriptscriptstyle{(0)}}{g}_{ij}),
\end{eqnarray} can be determined completely by PBH transformation \cite{Theisen,Theisen1}. Of course, one can also derive eq.(\ref{g1}) by using equations of motion \cite{Henningson,Nojiri}. The key point is that all higher derivative gravity theories give the same $\overset{\scriptscriptstyle{(1)}}{g}_{ij}$ eq.(\ref{g1}), due to the symmetry near the AdS boundary \cite{Theisen}. Unlike $\overset{\scriptscriptstyle{(1)}}{g}_{ij}$, The higher order terms $\overset{\scriptscriptstyle{(2)}}{g}_{ij},\overset{\scriptscriptstyle{(3)}}{g}_{ij}...$ are constrained by equations of motion.  We have
\bea\label{g2}
\gs_{i j} &=& k_1\,  C_{m n k l}C^{m n k l}\gz_{i j} +
k_2\, C_{i k l m}C_{j}^{~~k l m} \nonumber \\
&& + \frac{1}{d-4}\bigg[\frac{1}{8(d-1)}\nabla_i\nabla_jR - \frac{1}{4(d-2)}\Box R_{i j}+
\frac{1}{8(d-1)(d-2)}\Box R \gz_{i j} \nonumber\\
&&-\frac{1}{2(d-2)}R^{k l}R_{i k j l} + \frac{d-4}{2(d-2)^2}R_{i}^{~k}R_{j k}+
\frac{1}{(d-1)(d-2)^2}RR_{i j}\nonumber \\
&&+\frac{1}{4(d-2)^2}R^{k l}R_{k l}\gz_{i j}
-\frac{3d}{16(d-1)^2(d-2)^2}R^2 \gz_{i j}\bigg]\,,
\eea
where we have ignored $\scriptscriptstyle{(0)}$ in the above equation and $k_1, k_2$ depend on the action, or equivalently, equations of motion. For example, we have
\begin{eqnarray}\label{k1k2}
	k_1=\frac{3}{80}(5\lambda_1+14\lambda_2) ,\ \ k_2=\frac{3}{4}(\lambda_1-4\lambda_2)
\end{eqnarray}
for the action eq.(\ref{smartaction}).

Using the universal identity eq.(\ref{g1}), we can derive some useful formulas \cite{Miao1}
\begin{eqnarray}\label{order2}
	&&\tilde{R}\sim o(\rho^2),\ \ \tilde{R}_{ij}\sim o(\rho),\
	\ \ \tilde{R}_{i\rho}\sim o(\rho), \ \ \tilde{R}_{\rho\rho}\sim o(1)\nonumber\\
	&&\tilde{R}_{i\rho j \rho}\sim o(\frac{1}{\rho}),\ \ \tilde{R}_{\rho ijk}\sim o(\frac{1}{\rho})\nonumber\\
	&&\tilde{R}_{ijkl}=\frac{C_{ijkl}}{\rho}.
\end{eqnarray}
where $\tilde{R}$ eq.(\ref{diffcurvature}) is the difference between the curvature and a background-curvature. The above equations imply $O(\tilde{R}^n)$ are at least of order $O(\rho^n)$. For example, we have 
\begin{eqnarray}\label{Rn}
	\tilde{R}_{\mu\nu\rho\sigma}\tilde{R}^{\mu\nu\rho\sigma}&=&\rho^2 C_{ijkl}C^{ijkl}+O(\rho^3)\\
	\tilde{R}^{\mu\nu}_{\ \ \rho\sigma}\tilde{R}^{\rho\sigma}_{\ \ \alpha\beta}\tilde{R}^{\alpha\beta}_{\ \ \mu\nu}&=&\rho^3 C^{ij}_{\ \ mn} C^{mn}_{\ \ \ kl} C^{kl}_{\ \ ij}+O(\rho^4)\\
\end{eqnarray}
It should be stressed that the above results are on-shell, since we have used the on-shell condition eq.(\ref{g1}).

\subsubsection{Schwimmer-Theisen approach}

Denote the transverse space of the cones by $m$. The embedding of the $(d-1)$-dimensional submanifold $m$ into $(d+1)$-dimensional bulk is described by $X^{\mu}=X^{\mu}(\sigma^{\alpha})$, where $X^{\mu}=\{x^i, \rho\}$ are bulk coordinates and $\sigma^{\alpha}=\{y^{\hat{i}},\tau\}$ are coordinates on $m$. We choose a gauge
\begin{eqnarray}\label{gaugeLog1}
	\tau=\rho,\ \ h_{\tau \hat{i}}=0,
\end{eqnarray}
where $h_{\alpha\beta}$ is the induced metric on $m$.  Let us expand the embedding functions as
\begin{eqnarray}\label{embeddingfunctions}
	X^i(\tau,y^{\hat{j}})=\overset{\scriptscriptstyle{(0)}}{X^i}(y^{\hat{j}})+\overset{\scriptscriptstyle{(1)}}{X^i}(y^{\hat{j}})\tau+\overset{\scriptscriptstyle{(2)}}{X^i}(y^{\hat{j}})\tau^2+...
\end{eqnarray}
Diffeomorphism preserving the Fefferman-Graham gauge (\ref{FGgauge}) and above gauge (\ref{gaugeLog1}) uniquely fixes a transformation rule of the embedding functions $X^{\mu}(y^{\hat{i}},\tau)$ \cite{Theisen2}. From this transformation rule, we can identity $\overset{\scriptscriptstyle{(1)}}{X^i}(y^{\hat{j}})$ with the extrinsic curvature of the entangling surface $\Sigma$ on the AdS boundary
\begin{eqnarray}\label{rules1}
	\overset{\scriptscriptstyle{(1)}}{X^i}(y^{\hat{j}})=\frac{1}{2(d-2)}k^i(y^{\hat{j}}),
\end{eqnarray}
Similar to $\overset{\scriptscriptstyle{(1)}}{g}_{ij}$ eq.(\ref{g1}), $\overset{\scriptscriptstyle{(1)}}{X^i}$ is fixed by the symmetry and thus is universal \cite{Theisen2}. Of course, one can also get eq.(\ref{rules1}) by calculating the extremum of the entropy functional \footnote{Please refer to the Appendix for the details.}. Actually, as we shall show below, eq.(\ref{rules1}) is the perturbative counterpart of the extremal-area-surface condition $K^{i}=0$. Similar to $\overset{\scriptscriptstyle{(n)}}{g}_{ij}$, $\overset{\scriptscriptstyle{(n)}}{X^i}$ with $n\ge 2$ are non-universal and depend on the gravitational theories. Fortunately, we do not need $\overset{\scriptscriptstyle{(n)}}{X^i}$ with $n\ge 2$ for the derivations of the universal terms of EE in six-dimensional space-time.

From eqs.(\ref{embeddingfunctions},\ref{rules1}), we can derive the induced metric on $m$ as
\begin{eqnarray}\label{inducedhabA}
	h_{\tau\tau}&=&\partial_{\tau}X^{\mu}\partial_{\tau}X^{\nu}\hat{G}_{\mu\nu}=\frac{1}{4\tau^2}\Big{(}1+\frac{1}{(d-2)^2}\,
	k^ik^j\overset{\scriptscriptstyle{(0)}}{g}_{ij}\,\tau+\cdots\Big{)},\\
	h_{\hat{i}\hat{j}}&=&\frac{1}{\rho}\partial_{\hat{i}}X^{i}\partial_{\hat{j}}X^{j}g_{ij}=\frac{1}{ \tau}\Big{(}\overset{\scriptscriptstyle{(0)}}{h}_{\hat{i}\hat{j}}+\overset{\scriptscriptstyle{(1)}}{h}_{\hat{i}\hat{j}}\,\tau+...\Big{)},\label{inducedhabB}
\end{eqnarray}
with
\begin{eqnarray}\label{inducedhab2}
	\overset{\scriptscriptstyle{(0)}}{h}_{\hat{i}\hat{j}}=\partial_{\hat{i}} \overset{\scriptscriptstyle{(0)}}{X^i}\partial_{\hat{j}} \overset{\scriptscriptstyle{(0)}}{X^j}\ \overset{\scriptscriptstyle{(0)}}{g}_{ij},\ \ \ \overset{\scriptscriptstyle{(1)}}{h}_{\hat{i}\hat{j}}=\overset{\scriptscriptstyle{(1)}}{g}_{\hat{i}\hat{j}}-\frac{1}{d-2}k^ik^j_{\hat{i}\hat{j}} \overset{\scriptscriptstyle{(0)}}{g}_{ij}.
\end{eqnarray}
Thus, we have
\begin{eqnarray}\label{Deth}
	\sqrt{h}=\sqrt{\overset{\scriptscriptstyle{(0)}}{h}}\frac{1}{2\rho^{\frac{d}{2}}}\big( 1+\frac{\rho}{2}(\overset{\scriptscriptstyle{(0)}}{h}{}^{\hat{i}\hat{j}}\overset{\scriptscriptstyle{(1)}}{g}_{\hat{i}\hat{j}}-\frac{d-3}{(d-2)^2}\ k^ik^j \overset{\scriptscriptstyle{(0)}}{g}_{ij})+...\big)
\end{eqnarray}
Using eq.(\ref{embeddingfunctions},\ref{rules1}), we can also derive the extrinsic curvature $K$ of $m$ as
\begin{eqnarray}\label{hatK}
	K^i_{\hat{i}\hat{j}}=(k^i_{\hat{i}\hat{j}}-\frac{k^i}{d-2}\overset{\scriptscriptstyle{(0)}}{h}_{\hat{i}\hat{j}})+...
\end{eqnarray}
One can check that all the other components of $K^{\mu}_{\alpha\beta}$ are higher order terms irrelevant to the logarithmic terms. Notice that the leading term of $K^i_{\hat{i}\hat{j}}$ eq.(\ref{hatK}) is traceless, which means eq.(\ref{rules1}) yields the extremal-area-surface condition $K^{i}=0$ perturbatively. 

Now let us list some useful formulas in the notation of \cite{Dong}
\begin{eqnarray}\label{KKDong}
	K_{aij}=\frac{\bar{k}_{aij}}{\sqrt{\rho}}+...,\  
	K_a\sim \rho^{3/2},\ \ 
	K_{aij}K^{aij}=\rho \bar{k}_{aij}\bar{k}^{aij}+...
\end{eqnarray}
where $\bar{k}_{aij}$ denotes the traceless part of the extrinsic curvature. 

To derive the universal logarithmic terms of EE, we need to select the $\frac{1}{\rho}$ terms in the integrand of the entropy functional. From eqs.(\ref{Deth},\ref{KKDong}), it is easy to find that only following combinations of the extrinsic curvatures contribute to logarithmic terms of EE
\begin{eqnarray}\label{4dCFTlogterms}
	\sqrt{h}, \ \ \ \sqrt{h} tr K^2, \ \  \ \  \ \  \ \ \text{for}\ d=4
\end{eqnarray}
\begin{eqnarray}\label{KKDong1}
	\sqrt{h}, \ \ \ \sqrt{h} tr K^2,\ \ \sqrt{h} tr K^4,\ \  \sqrt{h} (tr K^2)^2,\  \ \ \text{for}\ d=6
\end{eqnarray}
Using eq.(\ref{KKDong}), we obtain some very useful formulas 
\begin{eqnarray}\label{CFTlogterms4d}
	\sqrt{h} tr K^2= \frac{1}{2\rho}\sqrt{\overset{\scriptscriptstyle{(0)}}{h}}\ tr \bar{k}^2+... \ \text{for} \ d=4.
\end{eqnarray}
\begin{eqnarray}\label{CFTlogterms6d1}
	\sqrt{h} tr K^4&=& \frac{1}{2\rho}\sqrt{\overset{\scriptscriptstyle{(0)}}{h}}\ tr \bar{k}^4+... \ \text{for}\ d=6.\\
	\sqrt{h} (tr K^2)^2&=& \frac{1}{2\rho}\sqrt{\overset{\scriptscriptstyle{(0)}}{h}}\ tr (\bar{k}^2)^2+... \ \text{for}\ d=6.\label{CFTlogterms6d2}
\end{eqnarray}
Notice that we have used the universal identity eq.(\ref{rules1}) in the above equations. Since eq.(\ref{rules1}) is the perturbative counterpart of the extremal-area-surface condition, therefore  eqs.(\ref{CFTlogterms4d},\ref{CFTlogterms6d1},\ref{CFTlogterms6d2}) are on-shell.

\subsection{Example: higher curvature gravity}

Now we take the general higher curvature gravity to illustrate how to derive the universal terms of EE in the approach of \cite{Theisen2}. For simplicity, we focus on 4d CFTs in this subsection and leave the study of 6d CFTs to the next section. 

We use the 'background field approach' introduced in \cite{Miao1}. This method together with \cite{Theisen2,Theisen} are very useful tools to derive the holographic Weyl anomaly and universal terms of EE \cite{Miao1}. Firstly, we define a 'background-curvature' (we set the AdS radius $l=1$)
\begin{eqnarray}\label{backcurvature}
	\bar{R}_{\mu\nu\sigma\rho}=\hat{G}_{\mu\rho}\hat{G}_{\nu\sigma}-\hat{G}_{\mu\sigma}\hat{G}_{\nu\rho}
\end{eqnarray}
and denote the difference between the curvature and the  'background-curvature' by
\begin{eqnarray}\label{diffcurvature}
	\tilde{R}_{\mu\nu\sigma\rho}=\hat{R}_{\mu\nu\sigma\rho}-\bar{R}_{\mu\nu\sigma\rho}.
\end{eqnarray}
Then we expand the action around this 'background-curvature' and get 
\begin{eqnarray}\label{GHCaction0}
	I&=&\frac{1}{16\pi G}\int d^{d+1}X\sqrt{\hat{G}} f(\hat{R}_{\mu\nu\sigma\rho})\nonumber\\
	&=&\frac{1}{16\pi G}\int d^{d+1}X\sqrt{\hat{G}}\big[ f_0+ c^{(1)}_1 \tilde{R} +( c^{(2)}_1\tilde{R}_{\mu\nu\sigma\rho}\tilde{R}^{\mu\nu\sigma\rho}+ c^{(2)}_2\tilde{R}_{\mu\nu}\tilde{R}^{\mu\nu}+ c^{(2)}_3\tilde{R}^2)+O(\tilde{R}^3) \big]\nonumber\\
\end{eqnarray}
where $f_0=f(\bar{R}_{\mu\nu\sigma\rho})=f(\hat{R}_{\mu\nu\sigma\rho})|_{AdS}$ is the Lagrangian for pure AdS, $c^{(n)}_i$ are some constants determined by the action. We require that the higher derivative gravity has an asymptotic AdS solution. This would impose a condition $ c^{(1)}_1=-f_0/2d$ \cite{Miao1}. Using this condition, we can rewrite the action (\ref{GHCaction0}) as
\begin{eqnarray}\label{GHCaction1}
	I=\frac{1}{16\pi G}\int d^{d+1}X\sqrt{\hat{G}}\big[-\frac{f_0}{2d}(\hat{R}+d^2-d)+( c^{(2)}_1\tilde{R}_{\mu\nu\sigma\rho}\tilde{R}^{\mu\nu\sigma\rho}+ c^{(2)}_2\tilde{R}_{\mu\nu}\tilde{R}^{\mu\nu}+ c^{(2)}_3\tilde{R}^2)+O(\tilde{R}^3) \big]\nonumber\\
\end{eqnarray}

Now we focus on the case of $d=4$. Applying the entropy formula (\ref{HEEDong}), we can derive HEE in asymptoticlly $AdS_5$ (\ref{FGgauge}) as
\begin{eqnarray}\label{HEE4dCFT}
	S=\frac{1}{4 G}\int d^{3}y \sqrt{h}\big{[} -\frac{f_0}{8}+2c^{(2)}_3 \tilde{R}+ c^{(2)}_2(\tilde{R}^a_a-\frac{1}{2}K_aK^a)+2 c^{(2)}_1(\tilde{R}^{ab}_{\ \ ab}-K_{aij}K^{aij}) + O(K^4, \tilde{R}^2, \tilde{R} K^2)  \big{]}\nonumber\\
\end{eqnarray} 
where  $a,b \in \{z,\bar{z}\}$ denotes orthogonal directions to the bulk entangling surface $m$. 
From eqs.(\ref{FGgauge},\ref{order2},\ref{KKDong}), it is not difficult to find that $\tilde{R}\sim \tilde{R}^a_a\sim O(K^4, \tilde{R}^2, \tilde{R} K^2)\sim O(\rho^2)$ are some higher order terms irrelevant to the logarithmic terms. Notice that, for 4d CFTs, we can ignore the splitting problem, since it only affects the entropy at order $O(K^4)$. Recall that we have
\begin{eqnarray}\label{4dlogCFTformula}
	&& \int_{\delta^2} \frac{d\rho}{2\rho} =\log (l/\delta),\ \sqrt{h} K_aK^a= \sqrt{h}(tr K)^2\sim O(\rho)\\
	&&\sqrt{h} K_{aij}K^{aij}=\sqrt{h} tr K^2= \frac{1}{2\rho}\sqrt{\overset{\scriptscriptstyle{(0)}}{h}}\ tr \bar{k}^2+ O(1)\label{4dlogCFTformula1}\\
	&&\sqrt{h} \tilde{R}^{ab}_{\ \ ab}= \frac{1}{2\rho}\sqrt{\overset{\scriptscriptstyle{(0)}}{h}}C^{ab}_{\ \ ab}+ O(1)= \frac{1}{2\rho}\sqrt{\overset{\scriptscriptstyle{(0)}}{h}}\overset{\scriptscriptstyle{(0)}}{h}{}^{\hat{i}\hat{j}}\overset{\scriptscriptstyle{(0)}}{h}{}^{\hat{k}\hat{l}}C_{\hat{i}\hat{k}\hat{j}\hat{l}}+ O(1)\label{4dlogCFTformula2}\\
	&& \sqrt{h}=\sqrt{\overset{\scriptscriptstyle{(0)}}{h}}\frac{1}{2\rho^{2}}\big( 1+\frac{\rho}{2}(\overset{\scriptscriptstyle{(0)}}{h}{}^{\hat{i}\hat{j}}\overset{\scriptscriptstyle{(1)}}{g}_{\hat{i}\hat{j}}-\frac{1}{4}\ k^ik^j \overset{\scriptscriptstyle{(0)}}{g}_{ij})+...\big)\nonumber\\
	&&\ \ \ \ \ =\sqrt{\overset{\scriptscriptstyle{(0)}}{h}}\frac{1}{2\rho^{2}}\big( 1+\frac{\rho}{4}(\overset{\scriptscriptstyle{(0)}}{h}{}^{\hat{i}\hat{j}}\overset{\scriptscriptstyle{(0)}}{h}{}^{\hat{k}\hat{l}}C_{\hat{i}\hat{k}\hat{j}\hat{l}}-tr \bar{k}^2-R_{\Sigma})+...\big)\label{4dlogCFTformula3}
\end{eqnarray} 
where $R_{\Sigma}$ is the intrinsic curvature scalar on the boundary entangling surface $\Sigma$ and we have used the variant of the Gauss-Codazzi equation in the above calculations.
\begin{eqnarray}\label{GaussCodazzi}
	2\overset{\scriptscriptstyle{(0)}}{h}{}^{\hat{i}\hat{j}}\overset{\scriptscriptstyle{(1)}}{g}_{\hat{i}\hat{j}}-\frac{1}{2}\ k^ik^j \overset{\scriptscriptstyle{(0)}}{g}_{ij}=\overset{\scriptscriptstyle{(0)}}{h}{}^{\hat{i}\hat{j}}\overset{\scriptscriptstyle{(0)}}{h}{}^{\hat{k}\hat{l}}C_{\hat{i}\hat{k}\hat{j}\hat{l}}-tr \bar{k}^2-R_{\Sigma}. 
\end{eqnarray}

Substituting eqs.(\ref{4dlogCFTformula}-\ref{4dlogCFTformula3}) into eq.(\ref{HEE4dCFT}), we can derive the universal logarithmic terms of EE as
\begin{eqnarray}\label{log4dCFT}
	S_{\text{log}}&=&\frac{1}{4 G}\int_{\delta^2} \frac{d\rho}{2\rho}\int_{\Sigma} d^{2}y \sqrt{\overset{\scriptscriptstyle{(0)}}{h}}\big{[} (-\frac{f_0}{32}+2c^{(2)}_1)(\overset{\scriptscriptstyle{(0)}}{h}{}^{\hat{i}\hat{j}}\overset{\scriptscriptstyle{(0)}}{h}{}^{\hat{k}\hat{l}}C_{\hat{i}\hat{k}\hat{j}\hat{l}}-tr \bar{k}^2) + \frac{f_0}{32} R_{\Sigma} \big{]}\nonumber\\
	&=&\frac{\log(l/\delta)}{2\pi}\int_{\Sigma} d^{2}y \sqrt{\overset{\scriptscriptstyle{(0)}}{h}}\big{[} c(\overset{\scriptscriptstyle{(0)}}{h}{}^{\hat{i}\hat{j}}\overset{\scriptscriptstyle{(0)}}{h}{}^{\hat{k}\hat{l}}C_{\hat{i}\hat{k}\hat{j}\hat{l}}-tr \bar{k}^2) + a R_{\Sigma} \big{]}
\end{eqnarray} 
where $a$ and $c$ are the central charges of 4d CFTs given by \cite{Miao1}
\begin{eqnarray}\label{centralcharge4dCFT}
	a=-\frac{f_0\pi}{64 G},\ \ c=-\frac{f_0\pi}{64 G}+ c^{(2)}_1 \frac{\pi}{G}
\end{eqnarray} 

Now we finish the derivations of the universal terms of EE for 4d CFTs dual to the general higher curvature gravity. Let us make some comments. Firstly, we have used the universal relations for  $\overset{\scriptscriptstyle{(1)}}{g}_{ij}$ eq.(\ref{g1}) and $\overset{\scriptscriptstyle{(1)}}{X^i}$ eq.(\ref{rules1}) in the above calculations. It seems that eq.(\ref{g1}) and eq.(\ref{rules1}) are off-shell, since they are obtained by making no use of equations of motion \cite{Theisen2}. However, this is not the case. Actually  eqs.(\ref{g1},\ref{rules1}) can also be derived from equations of motion and the extremal entropy condition (see Appendix. A), therefore the approach of \cite{Theisen2} is indeed on-shell. Secondly, expanding the action around the background-curvature eq.(\ref{backcurvature}) are quite useful for the derivations of universal terms of EE. At the leading order one can always directly replace $\tilde{R}^{ij}_{\ \ kl}$  ($K_{a\ j}^{\ i}$) by $\rho\ C^{ij}_{\ \ kl}$  ($\sqrt{\rho}\ \bar{k}_{a\ j}^{\ i}$), which can simplify the calculations greatly. For example, we have $\sqrt{h} tr K^2\to \log(l/\delta)\sqrt{\overset{\scriptscriptstyle{(0)}}{h}}\ tr \bar{k}^2$ for $d=4$, and $\sqrt{h} tr (\tilde{R} K^2)\to  \log(l/\delta)\sqrt{\overset{\scriptscriptstyle{(0)}}{h}}\ tr (C \bar{k}^2)$ for $d=6.$ Thirdly, the splitting problem does not matter for the universal terms of EE in four dimensions. However, it does matter for the case of 6d CFTs. We investigate this problem carefully in the next section.

\section{Holographic method}

In this section, we derive the universal logarithmic terms of EE for 6d CFTs by using the holographic method introduced in Sect. 3. We firstly derive the results from a smart-constructed bulk action and then prove that the general action produces the same results.

\subsection{Logarithmic terms of EE from a smart-constructed action}

For the curvature-squared gravity and Lovelock gravity, the splitting problem does not matter. However, the central charges of the corresponding CFTs are not independent but constrained by $B_3=\frac{B_2-\frac{B_1}{2}}{3}$. To cover the general CFTs, we must consider at least one cubic curvature term. Below we construct two special cubic curvature terms $M_1$ and $M_2$, which are designed to correspond to the conformal invariants $I_1$ and $I_2$ eq.(\ref{trace6x}) on the boundary, respectively.  We use these smart-constructed cubic curvature terms to derive universal terms of EE for 6d CFTs. It turns out that they help quite a lot to simplify the calculations. 

Consider the following action
\begin{eqnarray}\label{smartaction}
	S=\int
	d^{7}X\sqrt{-\hat{G}}(\hat{R}+30+\lambda_1 M_1+\lambda_2 M_2 )
\end{eqnarray}
where we have set the AdS radius $l=1$ and $M_1, M_2$ are constructed as
\begin{eqnarray}\label{Mi}
	&&M_1=\tilde{R}_{\mu\nu\rho\sigma}\tilde{R}^{\mu\alpha\beta\sigma }\tilde{R}^{\nu}_{\ \alpha\beta}{}^{\rho},\ \ \ \ \ M_2=\tilde{R}_{\mu\nu}^{\ \ \rho\sigma}\tilde{R}_{\rho\sigma}^{\ \ \alpha\beta}\tilde{R}_{\alpha\beta}^{\ \ \mu\nu}.
\end{eqnarray}
Recall that $\tilde{R}$ are defined by
\begin{eqnarray}\label{hatR}
	&&\tilde{R}_{\mu\nu\rho\sigma}=\hat{R}_{\mu\nu\rho\sigma}+(\hat{G}_{\mu\rho}\hat{G}_{\nu\sigma}-\hat{G}_{\mu\sigma}\hat{G}_{\nu\rho}),\nonumber\\
	&&\tilde{R}_{\mu\nu}=\hat{R}_{\mu\nu}+6 \hat{G}_{\mu\nu},\nonumber\\
	&&\tilde{R}=\hat{R}+42.
\end{eqnarray}
It should be mentioned that $M_i$ ($i=1,2$) can be regarded as the bulk counterparts to the conformal invariants $I_i$ eq.(\ref{trace6x}). They only contribute to the holographic Weyl anomaly with respect to $I_i$ ($i=1,2$).  According to \cite{Miao1}, the holographic Weyl anomaly for the above action is 
\be
\langle\,T^i{}_i\,\rangle=\sum_{n=1}^3 B_n\, I_n + 2 A \, E_6,
\ee
with the central charges given by
\begin{eqnarray}\label{centralcharges}
	&A&=\pi^3,\nonumber\\
	&B_1&=-\frac{1}{16}+\lambda_1,\nonumber\\
	&B_2&=-\frac{1}{64}+\lambda_2,\nonumber\\
	&B_3&=\frac{1}{192}.
\end{eqnarray}
It is expected that the universal terms of EE for 6d CTFs takes the following form
\begin{eqnarray}\label{logEEform}
	S_\mt{EE} = \log(\ell/\delta)\, \int_{\Sigma} d^{4}x\sqrt{h_0}\,\left[\,
	2\pi\sum_{n=1}^3 B_n F_n +2\,A\,
	E_{4}\,\right]
\end{eqnarray}
where $F_n$ are conformal invariants need to be determined and $E_4$ is the Euler density. From eqs.(\ref{smartaction},\ref{centralcharges}), it is clear that we can use HEE of $M_1$ and $M_2$ to derive $F_1$ and $F_2$, respectively. Knowing $F_1$ and $F_2$, one can use HEE of Einstein gravity to obtain $F_3$.

\subsubsection{$F_1$ and $F_2$}

Now let us start to derive the universal terms of EE by applying the approach introduced in Sect. 3.2. For convenience of the readers, we recall some useful formulas
\begin{eqnarray}\label{6dlogCFTformula}
	&& \int_{\delta^2} \frac{d\rho}{2\rho} =\log (l/\delta),\ \ \ \sqrt{h}(tr K)^4\sim O(\rho^3)\\
	&& \sqrt{h}(tr K)^2(tr K^2)\sim O(\rho),\ \sqrt{h}(tr K) (tr K^3)\sim O(1)\label{6dlogCFTformula1}\\
	&&\sqrt{h} tr K^4= \frac{1}{2\rho}\sqrt{\overset{\scriptscriptstyle{(0)}}{h}}\ tr \bar{k}^4+ O(1), \ \sqrt{h} (tr K^2)^2= \frac{1}{2\rho}\sqrt{\overset{\scriptscriptstyle{(0)}}{h}}\ (tr \bar{k}^2)^2+ O(1)\label{6dlogCFTformula2}\\
	&&\sqrt{h} Tr (\hat{\epsilon}\hat{\epsilon}\tilde{R}\tilde{R})= \frac{1}{2\rho}\sqrt{\overset{\scriptscriptstyle{(0)}}{h}}Tr (\tilde{\epsilon}\tilde{\epsilon}CC)+ O(1),\ \sqrt{h} Tr (K^2\tilde{R})= \frac{1}{2\rho}\sqrt{\overset{\scriptscriptstyle{(0)}}{h}}Tr (\bar{k}^2C)+ O(1)\nonumber\\\label{6dlogCFTformula3}
\end{eqnarray}
where $\hat{\epsilon}_{\mu\nu}$ and $\tilde{\epsilon}_{ij}$ are the two-dimensional volume form in the space transverse to the entangling surfaces in the bulk and on the boundary, respectively.

Let us firstly discuss the logarithmic terms from the Wald entropy of the action eq.(\ref{smartaction}). After some calculations, we get
\begin{eqnarray}\label{Waldsmartaction}
	S_{\text{Wald}}&=&2\pi\int d\rho d^4y \sqrt{h}\big{[}2+3\lambda_1\hat{\epsilon}^{\mu\nu}\hat{\epsilon}_{\rho\sigma}\tilde{R}_{\mu}^{\ \alpha\beta\sigma }\tilde{R}_{\nu\alpha\beta}{}^{\rho}+3\lambda_2 \hat{\epsilon}^{\mu\nu}\hat{\epsilon}_{\rho\sigma}\tilde{R}^{\rho\sigma\alpha\beta}\tilde{R}_{\alpha\beta\mu\nu} \nonumber\\
	&=&2\pi\int d\rho d^4y \sqrt{h}\big{[}2+\rho^2(3\lambda_1\tilde{\epsilon}^{ij}\tilde{\epsilon}_{kl}C_{i}^{\ mnl }C_{jmn}{}^{k}+3\lambda_2\tilde{\epsilon}^{ij}\tilde{\epsilon}_{kl}C^{kl}_{\ \ mn }C^{mn}_{\ \ \ \ ij}\big{]}+\text{irrelevant terms}\nonumber\\
	&=&S_{E}+2\pi\int d\rho d^4y\frac{ \sqrt{h_0}}{2\rho}\big{[} 3\lambda_1\tilde{\epsilon}^{ij}\tilde{\epsilon}_{kl}C_{i}^{\ mnl }C_{jmn}{}^{k}+3\lambda_2\tilde{\epsilon}^{ij}\tilde{\epsilon}_{kl}C^{kl}_{\ \ mn }C^{mn}_{\ \ \ \ ij}+(4k_1+k_2) C_{ijkl}C^{ijkl}\nonumber\\
	&&- k_2   g^{\perp}_{ij}C^{i}_{\ klm}C^{jklm} \big{]}+ \text{irrelevant terms}\nonumber\\
	&=&S_{E}+2\pi \log(\ell/\delta) \int_{\Sigma} d^{4}y\sqrt{h_0} \big{[}3\lambda_1 (C^{j m n k}\, C_{m~~n}^{~~i
		l}\,\tilde\veps_{i j}\,\tilde\veps_{k l} - \frac{1}{4}\,
	C^{iklm}\,C^{j}_{~kl m }\,\tilde{g}^{\perp}_{i j} + \frac{1}{20}\,
	C^{ijkl}\,C_{ijkl})\nonumber\\
	&&+3\lambda_2(C^{k l m n}\, C_{m n}^{~~~ i j}\,
	\tilde\veps_{i j}\,\tilde\veps_{k l} -C^{iklm}\,C^{j}_{~kl m
	}\,\tilde{g}^{\perp}_{i j} + \frac{1}{5}\, C^{ijkl}\,C_{ijkl})\big{]},
\end{eqnarray}
where $S_E$ is the universal terms of EE for Einstein gravity. We leave the derivation of $S_E$ to the next subsection. Let us discuss the above calculations briefly. The $R^3$ terms in action eq.(\ref{smartaction}) gives two kinds of contributions. The first kind of contributions come from their Wald entropy, such as the $C^2$ terms in the second and third lines of eq.(\ref{Waldsmartaction}). The second kind of contributions are due to their non-trivial corrections of $\overset{\scriptscriptstyle{(2)}}{g}_{ij}$ eq.(\ref{g2}) and $\overset{\scriptscriptstyle{(2)}}{X^i}$ eq.(\ref{embeddingfunctions}) in $\sqrt{h}$. The $k_1, k_2$ terms in the third and fourth lines of eq.(\ref{Waldsmartaction}) come from corrections of $\overset{\scriptscriptstyle{(2)}}{g}_{ij}$. Note that $\sqrt{h}$ contains only the linear term of $\overset{\scriptscriptstyle{(2)}}{X^i}$ in the relevant order $\frac{1}{\rho}$. According to equations of motion $\frac{\delta S_{\text{HEE}}}{\delta X^i}=0$, the linear terms of $\overset{\scriptscriptstyle{(2)}}{X^i}$ should vanish on-shell (at least for Einstein gravity). As we shall show in the next subsection, this is indeed the case.

From eqs.(\ref{centralcharges},\ref{logEEform},\ref{Waldsmartaction}), we can read out Wald-entropy-part of $F_1$ and $F_2$ as
\begin{eqnarray}\label{FW1}
	F_{W1}&=&3 (C^{j m n k}\, C_{m~~n}^{~~i
		l}\,\tilde\veps_{i j}\,\tilde\veps_{k l} - \frac{1}{4}\,
	C^{iklm}\,C^{j}_{~kl m }\,\tilde{g}^{\perp}_{i j} + \frac{1}{20}\,
	C^{ijkl}\,C_{ijkl})\\
	F_{W2}&=&3(C^{k l m n}\, C_{m n}^{~~~ i j}\,
	\tilde\veps_{i j}\,\tilde\veps_{k l} -C^{iklm}\,C^{j}_{~kl m
	}\,\tilde{g}^{\perp}_{i j} + \frac{1}{5}\, C^{ijkl}\,C_{ijkl}),\label{FW2}
\end{eqnarray}
which match the field theoretical results eqs.(\ref{CFTFW1},\ref{CFTFW2}) exactly.

Now let us go on to discuss the anomaly-like entropy for the action eq.(\ref{smartaction}). According to eqs.(\ref{6dlogCFTformula1},\ref{6dlogCFTformula2}), we only need to keep $tr K^4$ and $(tr K^2)^2$ among the $K^4$ terms. In other words, we can drop all terms including $K_{am}^{\ \ m}$. This helps a lot to simplify calculations. Note also that, as we have shown in Sect. 2, $Q_{0abij}\sim O(K^2), T_0\sim O(K^2)$.

For $M_1=\tilde{R}_{\mu\nu\rho\sigma}\tilde{R}^{\mu\alpha\beta\sigma }\tilde{R}^{\nu}_{\ \alpha\beta}{}^{\rho}$, applying the formula eq.(\ref{HEEDong}), we can derive the anomaly-like entropy as 
\begin{eqnarray}\label{AnomalyK8}
	S_{A_1}&=& \int d\rho d^4y \sqrt{h} \big{[}\ 24\pi K_{zij}K_{\bar{z}mn}\tilde{R}^{imjn}-12\pi K_{zij}K_{\bar{z}mn}(K_a^{in}K^{ajm}-K_a^{ij}K^{amn})\nonumber\\
	&&-96\pi K_{zil}K_{\bar{z}j}^{\ l}\tilde{R}_{z\bar{z}}^{\ \ ij}+48\pi K_{zil}K_{\bar{z}j}^{\ l}(K_{zjk}K_{\bar{z}i}^{\ \ k}-K_{zik}K_{\bar{z}j}^{\ \ k})\nonumber\\
	&&+96\pi K_{zij}K_{\bar{z}}^{\ ij}\tilde{R}_{z\bar{z}z\bar{z}}-48\pi K_{zij}K_{\bar{z}}^{\ ij}(-3T_0)\ \big{]}\\
	&=&\log(\ell/\delta) \int_{\Sigma} d^{4}y\sqrt{h_0} \big{[}24\pi \bar{k}_{zij}\bar{k}_{\bar{z}mn}C^{imjn}-12\pi \bar{k}_{zij}\bar{k}_{\bar{z}mn}(\bar{k}_a^{in}\bar{k}^{ajm}-\bar{k}_a^{ij}\bar{k}^{amn})\nonumber\\
	&&-96\pi \bar{k}_{zil}\bar{k}_{\bar{z}j}^{\ l}C_{z\bar{z}}^{\ \ ij}+48\pi \bar{k}_{zil}\bar{k}_{\bar{z}j}^{\ l}(\bar{k}_{zjk}\bar{k}_{\bar{z}i}^{\ \ k}-\bar{k}_{zik}\bar{k}_{\bar{z}j}^{\ \ k})\nonumber\\
	&&+96\pi \bar{k}_{zij}\bar{k}_{\bar{z}}^{\ ij}C_{z\bar{z}z\bar{z}}+24\pi(\bar{k}_{zij}\bar{k}_{\bar{z}}^{\ ij})^2\big{]}+...\label{LogAnomalyK8}
\end{eqnarray}
where $k_{aij}$ is the extrinsic curvature on the entangling surface $\Sigma$ and $\bar{k}_{aij}$ is the traceless part of $k_{aij}$. We have used eqs.(\ref{splittingEinstein1},\ref{6dlogCFTformula}-\ref{6dlogCFTformula3}) to derive eq.(\ref{LogAnomalyK8}) from eq.(\ref{AnomalyK8}).

Similarly for $M_2=\tilde{R}_{\mu\nu}^{\ \ \rho\sigma}\tilde{R}_{\rho\sigma}^{\ \ \alpha\beta}\tilde{R}_{\alpha\beta}^{\ \ \mu\nu}$, by using eqs.(\ref{HEEDong},\ref{splittingEinstein2},\ref{6dlogCFTformula}-\ref{6dlogCFTformula3}) we obtain
\begin{eqnarray}\label{AnomalyK7}
	S_{A_2}&=&\int d\rho d^4y \sqrt{h} \big{[}-384\pi K_{zl}^{\ i}K_{\bar{z}}^{\ jl}\tilde{R}_{zj\bar{z}i}+192\pi K_{zl}^{\ i}K_{\bar{z}}^{\ jl}(K_{zik}K_{\bar{z}j}^{\ \ k}-Q_{0z\bar{z}ij})\big{]}\\
	&=&\log(\ell/\delta) \int_{\Sigma} d^{4}y\sqrt{h_0} \big{[}-384\pi \bar{k}_{zl}^{\ i}\bar{k}_{\bar{z}}^{\ jl}C_{zj\bar{z}i}-192\pi \bar{k}_{zl}^{\ i}\bar{k}_{\bar{z}}^{\ jl}\bar{k}_{zjk}\bar{k}_{\bar{z}i}^{\ \ k}\big{]}+...\label{LogAnomalyK7}
\end{eqnarray}

Recall that $F_1$ and $F_2$ can be derived from the entropy of $M_1$ and $M_2$, respectively. Using eqs.(\ref{LogAnomalyK8},\ref{LogAnomalyK7}), we get the anomaly-like part of $F_1$ and $F_2$ as
\begin{eqnarray}\label{FA1}
	F_{A1}&=&\frac{S_{A_1}}{2\pi}=12 \bar{k}_{zij}\bar{k}_{\bar{z}mn}C^{imjn}-6 \bar{k}_{zij}\bar{k}_{\bar{z}mn}(\bar{k}_a^{in}\bar{k}^{ajm}-\bar{k}_a^{ij}\bar{k}^{amn})-48\bar{k}_{zil}\bar{k}_{\bar{z}j}^{\ l}C_{z\bar{z}}^{\ \ ij}\nonumber\\
	&&\ \ +24 \bar{k}_{zl}^{\ \ i}\bar{k}_{\bar{z}}^{\ lj}(\bar{k}_{zjk}\bar{k}_{\bar{z}i}^{\ \ k}-\bar{k}_{zik}\bar{k}_{\bar{z}j}^{\ \ k})+48 \bar{k}_{zij}\bar{k}_{\bar{z}}^{\ ij}C_{z\bar{z}z\bar{z}}+12(\bar{k}_{zij}\bar{k}_{\bar{z}}^{\ ij})^2\\
	F_{A2}&=&\frac{S_{A_2}}{2\pi}=-192\bar{k}_{zl}^{\ i}\bar{k}_{\bar{z}}^{\ jl}C_{zj\bar{z}i}-96\bar{k}_{zl}^{\ i}\bar{k}_{\bar{z}}^{\ jl}\bar{k}_{zjk}\bar{k}_{\bar{z}i}^{\ \ k}.\label{FA2}
\end{eqnarray}
Now we can obtain $F_1=F_{W1}+F_{A1}$ and $F_2=F_{W2}+F_{A2}$ from eqs.(\ref{FW1},\ref{FW2},\ref{FA1},\ref{FA2}). This is one of our main results. Let us make some comments. Firstly, we have used the splittings eqs.(\ref{splittingEinstein1},\ref{splittingEinstein2}), which implies that we require that our action has an asymptotically AdS solution.   Secondly, our results eqs.(\ref{FW1},\ref{FW2},\ref{FA1},\ref{FA2}) are consistent with those of \cite{Hung,Safdi}. We have shown above that our results agree with the field theoretical results eqs.(\ref{CFTFW1},\ref{CFTFW2}) when the extrinsic curvature vanishes\cite{Hung}. As for the case of non-zero extrinsic curvature, let us compare our results with \cite{Safdi}. In \cite{Safdi}, Safdi obtain the universal terms of EE for 6d CFTs with $B_3=\frac{B_2-\frac{B_1}{2}}{3}$ in flat space as eq.(\ref{6dlog}).  For simplicity Safdi takes vanishing extrinsic curvature in the time-like direction. In our notation, we have $k_{zij}=k_{\bar{z}ij}=\frac{1}{2}k_{ij}$.  Since now we do not know $F_3$, we set $B_3=0, B_1=2B_2$ for simplicity (We leave the derivation of $F_3$ to the next subsection). Note also that we have $C_{ijkl}=0$ in flat space. Take all the above simplifications into account, we derive 
\begin{eqnarray}\label{our6dlogflat}
	S_{\Sigma}|_{\log}=\log(\ell/\delta) \int_{\Sigma}  2A E_4+9\pi B_2[(tr\bar{k}^2)^2-2 tr \bar{k}^4]
\end{eqnarray}
which exactly agrees with the results of \cite{Safdi}. Thirdly,  our $\sqrt{h_0}F_1$ and $\sqrt{h_0}F_2$ are obviously conformally invariant. That is because, similar to $C_{ijkl}$, $\bar{k}_{aij}$ are conformal tensors. In other words, we have $g_{ij}\to e^{2\sigma}g_{ij}, C^i_{jkl}\to C^i_{jkl}$ and $\bar{k}_{aij}\to e^{\sigma} \bar{k}_{aij}$ under conformal transformations.  To end this section, we rewrite $F_1$ and $F_2$ in covariant expressions 
\begin{eqnarray}\label{F1}
	F_{1}&=&3 (C^{j m n k}\, C_{m~~n}^{~~i
		l}\,\tilde\veps_{i j}\,\tilde\veps_{k l} - \frac{1}{4}\,
	C^{iklm}\,C^{j}_{~kl m }\,\tilde{g}^{\perp}_{i j} + \frac{1}{20}\,
	C^{ijkl}\,C_{ijkl})\nonumber\\
	&&+3 \bar{k}^a_{\ ij}\bar{k}_{amn}C^{imjn}-\frac{3}{2} \bar{k}^b_{\ ij}\bar{k}_{bmn}(\bar{k}_a^{in}\bar{k}^{ajm}-\bar{k}_a^{ij}\bar{k}^{amn})\nonumber\\
	&&+3 \tilde\veps^{ab}\bar{k}_{a il}\bar{k}_{bj}^{\ l}\veps^{cd}C_{cd}^{\ \ ij}\tilde+3 \tilde\veps^{ab}\bar{k}_{a il}\bar{k}_{bj}^{\ l}\tilde\veps^{cd}\bar{k}_{c\ k}^{\ i}\bar{k}_{d}^{\ kj}\nonumber\\
	&&-\frac{3}{4} \bar{k}^a_{\ mn}\bar{k}_{a}^{\ mn}C^{ijkl}\,\tilde\veps_{i j}\,\tilde\veps_{k l} +\frac{3}{4}(\bar{k}^a_{\ ij}\bar{k}_{a}^{\ ij})^2
\end{eqnarray}

\begin{eqnarray}\label{F2}
	F_{2}&=&3(C^{k l m n}\, C_{m n}^{~~~ i j}\,
	\tilde\veps_{i j}\,\tilde\veps_{k l} -C^{iklm}\,C^{j}_{~kl m
	}\,\tilde{g}^{\perp}_{i j} + \frac{1}{5}\, C^{ijkl}\,C_{ijkl})\nonumber\\
	&&-12\bar{k}_{\ l}^{a\ (i}\bar{k}_{a}^{\ j)l}C_{minj}\tilde{g}^{\perp}{}^{mn} -6\bar{k}_{\ l}^{a\ (i}\bar{k}_{a}^{\ j)l}\bar{k}^b_{\ jk}\bar{k}_{bi}^{\ \ k}\nonumber\\
	&&+12\tilde\veps^{ab}\bar{k}_{a}^{\ il}\bar{k}_{bl}^{\ \ j}\tilde\veps^{cd}C_{cjdi} +6\tilde\veps^{ab}\bar{k}_{a}^{\ il}\bar{k}_{bl}^{\ \ j}\tilde\veps^{cd}\bar{k}_{c jk}\bar{k}_{d\ i}^{\ k}.
\end{eqnarray}

\subsubsection{$F_3$}

In this subsection, we derive the universal terms of EE for 6d CFTs dual to Einstein gravity. Using the result of this sub-section together with $F_1$ and $F_2$, we can derive $F_3$. 

Recall that the HEE of Einstein gravity is
\begin{eqnarray}\label{6dlogEinstein}
	S_{\text{HEE}}=4\pi \int d\rho d^4y \sqrt{h}
\end{eqnarray}
Applying the approach intruduced in Sect. 3.1.2, we have
\begin{eqnarray}\label{h00}
	h_{\rho\rho}&=&\frac{1}{4\rho^2}+\frac{1}{\rho} \overset{\scriptscriptstyle{(1)}}{X^j} \overset{\scriptscriptstyle{(1)}}{X^i} \overset{\scriptscriptstyle{(0)}}{g_{ij}} +(\overset{\scriptscriptstyle{(1)}}{X^i} \overset{\scriptscriptstyle{(1)}}{X^j} \overset{\scriptscriptstyle{(1)}}{g_{ij}}+\overset{\scriptscriptstyle{(1)}}{X^i} \overset{\scriptscriptstyle{(1)}}{X^j} \overset{\scriptscriptstyle{(1)}}{X^k} \partial_k\overset{\scriptscriptstyle{(0)}}{g_{ij}}  +4 \overset{\scriptscriptstyle{(2)}}{X^i} \overset{\scriptscriptstyle{(1)}}{X^j} \overset{\scriptscriptstyle{(0)}}{g_{ij}})\nonumber\\
	&=&\frac{1}{4\rho^2}[1+\frac{1}{16} \rho k^i k_i +  \rho^2(\frac{1}{16} k^i k^j\overset{\scriptscriptstyle{(1)}}{g_{ij}}+2\overset{\scriptscriptstyle{(2)}}{X^i} k^j \overset{\scriptscriptstyle{(0)}}{g_{ij}})].
\end{eqnarray}
Here we have used $\overset{\scriptscriptstyle{(1)}}{X^i}=\frac{1}{8}k^i$ eq.(\ref{rules1}) and the following ansatz of $\overset{\scriptscriptstyle{(0)}}{g_{ij}}$
\begin{eqnarray}\label{goij}
	\overset{\scriptscriptstyle{(0)}}{g_{ij}}dx^idx^j=dzd\bar{z}+T(\bar{z}dz-zd\bar{z})^2+2i V_{\hat{i}}(\bar{z}dz-zd\bar{z})dy^{\hat{i}}\nonumber\\
	+( g_{\hat{i}\hat{j}}+Q_{\hat{i}\hat{j}})dy^{\hat{i}}dy^{\hat{j}},
\end{eqnarray}
where $T, V, Q$ are given by
\begin{eqnarray}\label{TVQEinstein}
	&&T=\sum_{n=0}^{\infty}T_{a_1...a_n}x^{a_1}...x^{a_n},\ \ \ V_{\hat{i}}=\sum_{n=0}^{\infty} V_{ a_1...a_n \hat{i}}x^{a_1}...x^{a_n}=U_{\hat{i}}+...,\nonumber\\
	&&Q_{\hat{i}\hat{j}}=\sum_{n=1}^{\infty} Q_{a_1...a_n ij}x^{a_1}...x^{a_n}=-2 x^a k_{a\hat{i}\hat{j}}+x^ax^b Q_{ab\hat{i}\hat{j}}+...
\end{eqnarray}
Here $x^a$ denote $z, \bar{z}$ and $y^{\hat{i}}$ are coordinates on the four-dimensional entangling surface $\Sigma$. Using eq.(\ref{goij}), we have $\overset{\scriptscriptstyle{(1)}}{X^i} \overset{\scriptscriptstyle{(1)}}{X^j} \overset{\scriptscriptstyle{(1)}}{X^k} \partial_k\overset{\scriptscriptstyle{(0)}}{g_{ij}} \sim O(x^a)$ and thus can be ignored on the entangling surface $(x^a=0)$. It should be mentioned that,  by choosing suitable coordinates, we can alway write the metric in the form of eq.(\ref{goij}) \cite{Dong}.  Note also that the extrinsic curvature in this subsection (Schwimmer-Theisen notation \cite{Theisen2}) is different from the one of other sections (Dong's notation\cite{Dong}) by a minus sign.

Similarly for $h_{\hat{i}\hat{j}}$, we have
\begin{eqnarray}\label{hij}
	h_{\hat{i}\hat{j}}&=&\frac{1}{\rho}[\overset{\scriptscriptstyle{(0)}}{h}_{\hat{i}\hat{j}} +\rho(\overset{\scriptscriptstyle{(1)}}{g}_{\hat{i}\hat{j}}-\frac{1}{4}k^ak_{a\hat{i}\hat{j}})+\rho^2 \overset{\scriptscriptstyle{(2)}}{h}_{\hat{i}\hat{j}} ],
\end{eqnarray}
with $\overset{\scriptscriptstyle{(2)}}{h}_{\hat{i}\hat{j}}$ given by
\begin{eqnarray}\label{h2ij}
	\overset{\scriptscriptstyle{(2)}}{h}_{\hat{i}\hat{j}}&=&\partial_{\hat{i}}\overset{\scriptscriptstyle{(1)}}{X^m}\partial_{\hat{j}}\overset{\scriptscriptstyle{(1)}}{X^n}\overset{\scriptscriptstyle{(0)}}{g_{mn}}+\partial_{\hat{i}}\overset{\scriptscriptstyle{(2)}}{X^m}\partial_{\hat{j}}\overset{\scriptscriptstyle{(0)}}{X^n}\overset{\scriptscriptstyle{(0)}}{g_{mn}}+\partial_{\hat{i}}\overset{\scriptscriptstyle{(0)}}{X^m}\partial_{\hat{j}}\overset{\scriptscriptstyle{(2)}}{X^n}\overset{\scriptscriptstyle{(0)}}{g_{mn}}\nonumber\\
	&&+(\partial_{\hat{i}}\overset{\scriptscriptstyle{(1)}}{X^m}\partial_{\hat{j}}\overset{\scriptscriptstyle{(0)}}{X^n}+\partial_{\hat{i}}\overset{\scriptscriptstyle{(0)}}{X^m}\partial_{\hat{j}}\overset{\scriptscriptstyle{(1)}}{X^n})(\overset{\scriptscriptstyle{(1)}}{g_{mn}}+\overset{\scriptscriptstyle{(1)}}{X^k}\partial_k\overset{\scriptscriptstyle{(0)}}{g_{mn}})\nonumber\\
	&&+\partial_{\hat{i}}\overset{\scriptscriptstyle{(0)}}{X^m}\partial_{\hat{j}}\overset{\scriptscriptstyle{(0)}}{X^n}(\overset{\scriptscriptstyle{(2)}}{g_{mn}}+\overset{\scriptscriptstyle{(1)}}{X^k}\partial_k \overset{\scriptscriptstyle{(1)}}{g_{mn}}+\frac{\overset{\scriptscriptstyle{(1)}}{X^k}\overset{\scriptscriptstyle{(1)}}{X^l}}{2}\partial_k\partial_l\overset{\scriptscriptstyle{(0)}}{g_{mn}}+\overset{\scriptscriptstyle{(2)}}{X^k}\partial_k \overset{\scriptscriptstyle{(0)}}{g_{mn}})\nonumber\\
	&=&(\frac{1}{64}\partial_{\hat{i}}k^m\partial_{\hat{j}}k^n+\partial_{\hat{i}}\overset{\scriptscriptstyle{(2)}}{X^m}\partial_{\hat{j}}\overset{\scriptscriptstyle{(0)}}{X^n}+\partial_{\hat{i}}\overset{\scriptscriptstyle{(0)}}{X^m}\partial_{\hat{j}}\overset{\scriptscriptstyle{(2)}}{X^n})\overset{\scriptscriptstyle{(0)}}{g_{mn}}\nonumber\\
	&&+\frac{1}{8}(\partial_{\hat{i}}k^m\overset{\scriptscriptstyle{(1)}}{g}_{m\hat{j}}+\partial_{\hat{j}}k^m\overset{\scriptscriptstyle{(1)}}{g}_{m\hat{i}})+\frac{1}{32}\epsilon_{mn}(\partial_{\hat{i}}k^mk^n U_{\hat{j}}+\partial_{\hat{i}}k^mk^n U_{\hat{i}})\nonumber\\
	&&+\overset{\scriptscriptstyle{(2)}}{g}_{\hat{i}\hat{j}}+\frac{1}{8}k^a\partial_a\overset{\scriptscriptstyle{(1)}}{g}_{\hat{i}\hat{j}}+\frac{1}{64}k^ak^bQ_{ab\hat{i}\hat{j}}+\overset{\scriptscriptstyle{(2)}}{X^k}\partial_k \overset{\scriptscriptstyle{(0)}}{g_{mn}}
\end{eqnarray}
Let us try to simplify the above formula.
Focus on the $\overset{\scriptscriptstyle{(2)}}{X^m}$ terms which are relevant to the logarithmic  terms of EE, we have
\begin{eqnarray}\label{X2Einstein}
	S_{\overset{\scriptscriptstyle{(2)}}{X}}&=&4\pi\log(\ell/\delta)\int_{\Sigma} d^4y \sqrt{h_0}[\overset{\scriptscriptstyle{(2)}}{X^i} k^j \overset{\scriptscriptstyle{(0)}}{g_{ij}}+\overset{\scriptscriptstyle{(0)}}{h^{\hat{i}\hat{j}}}\partial_{\hat{i}}\overset{\scriptscriptstyle{(0)}}{X^m}\partial_{\hat{j}}\overset{\scriptscriptstyle{(2)}}{X^n}\overset{\scriptscriptstyle{(0)}}{g_{mn}}+\frac{1}{2}\overset{\scriptscriptstyle{(0)}}{h^{mn}}\overset{\scriptscriptstyle{(2)}}{X^k}\partial_k \overset{\scriptscriptstyle{(0)}}{g_{mn}}]\nonumber\\
	&=& 4\pi\log(\ell/\delta)\int_{\Sigma} d^4y \sqrt{h_0}\big{[}\overset{\scriptscriptstyle{(2)}}{X^i} \overset{\scriptscriptstyle{(0)}}{g_{ij}}\overset{\scriptscriptstyle{(0)}}{h^{\hat{m}\hat{n}}}(k^j_{\hat{m}\hat{n}}-\partial_{\hat{m}}\partial_{\hat{n}}\overset{\scriptscriptstyle{(0)}}{X^j}+\overset{\scriptscriptstyle{(0)}}{\gamma^{\hat{l}}}_{\hat{m}\hat{n}}\partial_{\hat{l}}\overset{\scriptscriptstyle{(0)}}{X^j}-\overset{\scriptscriptstyle{(0)}}{\Gamma^j}_{kl}\partial_{\hat{m}}\overset{\scriptscriptstyle{(0)}}{X^k}\partial_{\hat{n}}\overset{\scriptscriptstyle{(0)}}{X^l}\big{]}\nonumber\\
	&&+4\pi\log(\ell/\delta)\int_{\Sigma} d^4y\partial_{\hat{j}}(\sqrt{h_0}\overset{\scriptscriptstyle{(0)}}{h^{\hat{i}\hat{j}}}\partial_{\hat{i}}\overset{\scriptscriptstyle{(0)}}{X^m}\overset{\scriptscriptstyle{(2)}}{X^n}\overset{\scriptscriptstyle{(0)}}{g_{mn}})\nonumber\\
	&=& 4\pi\log(\ell/\delta)\int_{\Sigma} d^4y \sqrt{h_0} D_{\hat{i}}\overset{\scriptscriptstyle{(2)}}{X^{\hat{i}}}
\end{eqnarray}
where $\gamma^{\hat{l}}_{\hat{m}\hat{n}}$ and $D_{\hat{i}}$ are the Levi-Civita connection and covariant derivatives on the entangling surface $\Sigma$, respectively. In the above derivations, we have used the definition of the extrinsic curvature
\begin{eqnarray}\label{kimn} k^j_{\hat{m}\hat{n}}=\partial_{\hat{m}}\partial_{\hat{n}}\overset{\scriptscriptstyle{(0)}}{X^j}-\overset{\scriptscriptstyle{(0)}}{\gamma^{\hat{l}}}_{\hat{m}\hat{n}}\partial_{\hat{l}}\overset{\scriptscriptstyle{(0)}}{X^j}+\overset{\scriptscriptstyle{(0)}}{\Gamma^j}_{kl}\partial_{\hat{m}}\overset{\scriptscriptstyle{(0)}}{X^k}\partial_{\hat{n}}\overset{\scriptscriptstyle{(0)}}{X^l}.
\end{eqnarray}
Now it is clear that we can drop $\overset{\scriptscriptstyle{(2)}}{X}$ safely on closed entangling surfaces. Thus eq.(\ref{h2ij}) can be simplified as
\begin{eqnarray}\label{h2ij1}
	\overset{\scriptscriptstyle{(2)}}{h}_{\hat{i}\hat{j}}
	&=&\frac{1}{64}\partial_{\hat{i}}k^m\partial_{\hat{j}}k^n\overset{\scriptscriptstyle{(0)}}{g_{mn}}+\frac{1}{8}(\partial_{\hat{i}}k^m\overset{\scriptscriptstyle{(1)}}{g}_{m\hat{j}}+\partial_{\hat{j}}k^m\overset{\scriptscriptstyle{(1)}}{g}_{m\hat{i}})+\frac{1}{32}\epsilon_{mn}(\partial_{\hat{i}}k^mk^n U_{\hat{j}}+\partial_{\hat{i}}k^mk^n U_{\hat{i}})\nonumber\\
	&&+\overset{\scriptscriptstyle{(2)}}{g}_{\hat{i}\hat{j}}+\frac{1}{8}k^a\partial_a\overset{\scriptscriptstyle{(1)}}{g}_{\hat{i}\hat{j}}+\frac{1}{64}k^ak^bQ_{ab\hat{i}\hat{j}}\nonumber\\
	&=&\overset{\scriptscriptstyle{(0)}}{h}{}^{\ i}_{ \hat{i}}\overset{\scriptscriptstyle{(0)}}{h}{}^{\ j}_{\hat{j}}\big{[}\frac{1}{64}(\nabla_{i}k^m\nabla_{j}k^n\overset{\scriptscriptstyle{(0)}}{g_{mn}}-k^mk^nR_{minj})+\frac{1}{8}(\nabla_{i}k^m\overset{\scriptscriptstyle{(1)}}{g}_{mj}+\nabla_{j}k^m\overset{\scriptscriptstyle{(1)}}{g}_{mi}+k^m\nabla_m\overset{\scriptscriptstyle{(1)}}{g}_{ij})+\overset{\scriptscriptstyle{(2)}}{g}_{ij}\big{]},\nonumber\\
\end{eqnarray}
where $\nabla_i$ are the covariant derivatives with respect to $\overset{\scriptscriptstyle{(0)}}{g_{mn}}$. From eqs.(\ref{h00},\ref{h2ij1}), we can derive the logarithm term of EE for CFTs dual to Einstein gravity as
\begin{eqnarray}\label{SEinstein}
	S_{E}&=&\pi\log(\ell/\delta)\int_{\Sigma} d^4y \sqrt{h_0}[2\overset{\scriptscriptstyle{(2)}}{h^{\hat{i}}}_{\hat{i}}-\overset{\scriptscriptstyle{(1)}}{g}_{\hat{i}\hat{j}}\overset{\scriptscriptstyle{(1)}}{g}^{\hat{i}\hat{j}}+\frac{1}{2}(\overset{\scriptscriptstyle{(1)}}{g^{\hat{i}}}_{\hat{i}})^2+\frac{1}{2}k^ak_{a\hat{i}\hat{j}}\overset{\scriptscriptstyle{(1)}}{g}^{\hat{i}\hat{j}} -\frac{3}{16}k^ak_a\overset{\scriptscriptstyle{(1)}}{g^{\hat{i}}}_{\hat{i}}\nonumber\\
	&&\ \ \ \ \ \ \ \ \ \ \ \ \ \ \ \ \ \ \ \ \ \ \ \ \ \ \ \ \ \ +\frac{1}{8}k^ak^b\overset{\scriptscriptstyle{(1)}}{g_{ab}} -\frac{1}{16}k^ak_{a\hat{i}\hat{j}}k_bk^{b\hat{i}\hat{j}}+\frac{7}{512}(k^ak_a)^2  ].
\end{eqnarray}
The definitions of $\overset{\scriptscriptstyle{(1)}}{g},\overset{\scriptscriptstyle{(2)}}{g}$ can be  found in eqs.(\ref{g1},\ref{g2}) with $k_1=k_2=0$. After some complicated calculations, we find that eq.(\ref{SEinstein}) is conformally invariant up to some total derivative terms. This can be regarded as a check of eq.(\ref{SEinstein}). Please refer to Appendix B for the proof of the conformal invariance of  eq.(\ref{SEinstein}).  Using eq.(\ref{SEinstein}) together with $F_1$ and $F_2$ of sect. 2.2.1, we can derive $F_3$. 
\begin{eqnarray}\label{GeneralF3}
	F_3&=&-192\pi^2E_4+12F_1+3F_2\nonumber\\
	&&+192\big{[}\overset{\scriptscriptstyle{(2)}}{h^{\hat{i}}}_{\hat{i}}-\frac{1}{2}\overset{\scriptscriptstyle{(1)}}{g}_{\hat{i}\hat{j}}\overset{\scriptscriptstyle{(1)}}{g}^{\hat{i}\hat{j}}+\frac{1}{4}(\overset{\scriptscriptstyle{(1)}}{g^{\hat{i}}}_{\hat{i}})^2+\frac{1}{4}k^ak_{a\hat{i}\hat{j}}\overset{\scriptscriptstyle{(1)}}{g}^{\hat{i}\hat{j}} -\frac{3}{32}k^ak_a\overset{\scriptscriptstyle{(1)}}{g^{\hat{i}}}_{\hat{i}}\nonumber\\
	&&+\frac{1}{16}k^ak^b\overset{\scriptscriptstyle{(1)}}{g_{ab}} -\frac{1}{32}k^ak_{a\hat{i}\hat{j}}k_bk^{b\hat{i}\hat{j}}+\frac{7}{1024}(k^ak_a)^2 \big{]}
\end{eqnarray}
This is one of our main results. Now let us consider some special cases below.

Case I: $k_{aij}=0$,
\begin{eqnarray}\label{SEinstein1}
	S_{E}&=&\pi\log(\ell/\delta)\int_{\Sigma} d^4y \sqrt{h_0}[2\overset{\scriptscriptstyle{(2)}}{g^{\hat{i}}}_{\hat{i}}-\overset{\scriptscriptstyle{(1)}}{g}_{\hat{i}\hat{j}}\overset{\scriptscriptstyle{(1)}}{g}^{\hat{i}\hat{j}}+\frac{1}{2}(\overset{\scriptscriptstyle{(1)}}{g^{\hat{i}}}_{\hat{i}})^2]\nonumber\\
	&=&\log(\ell/\delta)\int_{\Sigma} d^4y \sqrt{h_0}[2\pi \sum_{n=1}^{3} B_n F_{W_n}+2A E_4+ B_3 \Delta S]\nonumber\\
\end{eqnarray}
where $F_{W_n}=\frac{\partial I_n}{\partial R^{i j}{}_{k l}}\,\tilde{\veps}^{i
	j}\,\tilde{\veps}_{k l}$ denote the Wald entropy eqs.(\ref{CFTFW1},\ref{CFTFW2},\ref{CFTFW3}). $B_n$ and $A$ are the central charges of CFTs dual to Einstein gravity, which can be found in eq.(\ref{centralcharges}) with $\lambda=0$. $\Delta S$ is the famous HMS mismatch \cite{Hung}, which was firstly found by Hung, Myers and Smolkin that the holographic universal terms of EE does not match the CFT ones even for entangling surface with zero extrinsic curvature. Recently, the authors of \cite{Miao2} find that HMS have ignored the anomaly-like entropy of $I_3$. Taking into account such contributions, the holographic and CFT results indeed match. After some tedious calculations, we derive $\Delta S$ as
\begin{eqnarray}\label{EinsteinDeltaS}
	\Delta S=-4\pi (&&C_{m n}{}^{r
		s }C^{m n k l} \tilde{g}^\perp_{s l}\tilde{g}^\perp_{r k} -C_{m n
		r}{}^s C^{m n r l}\tilde{g}^\perp_{s l} \nonumber\\
	&+& 2 C_{m}{}^n{}_r{}^s C^{mkrl} \tilde{g}^\perp_{n
		s}\tilde{g}^\perp_{k l} - 2C_{m}{}^n{}_r{}^s C^{mkrl}
	\tilde{g}^\perp_{n l}\tilde{g}^\perp_{k s}\nonumber\\
	&+&\frac{4}{3} \tilde{g}^\perp_{ij}\tilde{g}^\perp_{kl}\tilde{g}^\perp_{mn}\tilde{g}^\perp_{rs}C^{ikmr}C^{jlns}-\frac{4}{3} \tilde{g}^\perp_{ij}\tilde{g}^\perp_{kl}\tilde{g}^\perp_{mn}C^{ikm}_{\ \ \ \ s}C^{jlns})
\end{eqnarray}
Note that the first two lines of eq.(\ref{EinsteinDeltaS}) was derived by HMS \cite{Hung} under the conditions $k_{aij}=0$ and $R_{abci}=3\epsilon_{ab}V_{ci}=0$. If we drop the second condition, we get some new terms in the last line of eq.(\ref{EinsteinDeltaS}). Actually, these new terms are proportional to $R_{abci}R^{abci}$.

Case II: flat $\overset{\scriptscriptstyle{(0)}}{g_{ij}}$  and zero $\overset{\scriptscriptstyle{(1)}}{g_{ij}}=\overset{\scriptscriptstyle{(2)}}{g_{ij}}=0$. Note that this means the bulk spacetime is pure AdS.
\begin{eqnarray}\label{SEinstein2}
	S_{E}&=&\frac{\pi}{512}\log(\ell/\delta)\int_{\Sigma} d^4y \sqrt{h_0}[16 \partial_{\hat{i}}k^m \partial^{\hat{i}}k^n\overset{\scriptscriptstyle{(0)}}{g_{mn}}+7 (k^ak_a)^2-16 k^ak_{a\hat{i}\hat{j}}k_bk^{b\hat{i}\hat{j}}]
\end{eqnarray}
In the above derivations, we have used the flat condition $R_{aibj}=0$. For simplicity, we set $U_i=0$. This is also the case studied in \cite{Safdi}. Compare eq.(\ref{SEinstein2}) with 
\begin{eqnarray}\label{16}
	S_{\Sigma}|_{\log}&=&\log(\ell/\delta)\, \int_{\Sigma} d^{4}x\sqrt{h_0}\,\left[\,
	2\,A\,
	E_{4}+2\pi\sum_{n=1}^3 B_n F_n\,\right],
\end{eqnarray}
we can derive $F_3$ as
\begin{eqnarray}\label{EinsteinF3}
	F_3=\frac{3}{16}(16 \partial_{\hat{i}}k^m \partial^{\hat{i}}k^n\overset{\scriptscriptstyle{(0)}}{g_{mn}}+7 (k^ak_a)^2-16 k^ak_{a\hat{i}\hat{j}}k_bk^{b\hat{i}\hat{j}})-192\pi^2E_4+12F_1+3F_2
\end{eqnarray}
with $E_4$ and $F_n$ given by
\begin{eqnarray}\label{E4Fn}
	E_4&=&\frac{1}{128\pi^2}\delta^{i_1i_2i_3i_4}_{j_1j_2j_3j_4}R_{\Sigma}^{j_1j_2}{}_{i_1i_2}R_{\Sigma}^{j_3j_4}{}_{i_3i_4}=\frac{1}{32\pi^2}\delta^{i_1i_2i_3i_4}_{j_1j_2j_3j_4}k^{aj_1}_{\ \ \ i_1}k_{a}^{j_2}{}_{i_2}k^{bj_3}_{\ \ \ i_3}k_{b}^{j_4}{}_{i_4}\nonumber\\
	F_{1}&=&-\frac{3}{2} \bar{k}_{bij}\bar{k}^b_{\ mn}(\bar{k}_a^{in}\bar{k}^{ajm}-\bar{k}_a^{ij}\bar{k}^{amn})+3 \epsilon^{ab}\bar{k}_{ail}\bar{k}_{bj}^{\ l}\epsilon^{cd}\bar{k}_{ck}^{\ \ i}\bar{k}_{d}^{\ jk}+\frac{3}{4}(\bar{k}_{aij}\bar{k}^{aij})^2\nonumber\\
	F_{2}&=&-6\bar{k}_{al}^{\ \ i}\bar{k}^{alj}\bar{k}_{bjk}\bar{k}_{\ \ i}^{b\ \ k}+6\tilde\veps^{ab}\bar{k}_{a}^{\ il}\bar{k}_{bl}^{\  j}\tilde\veps^{cd}\bar{k}_{c jk}\bar{k}_{d\ i}^{\ k}
\end{eqnarray}
To derive $E_4$ in the above equation, we have used the `flat-space condition' $R^{\parallel}_{ijkl}=R_{\Sigma ijkl}-(k_{aik}k^a_{\ jl}-k_{ail}k^a_{\ jk})=0$. $F_1$ and $F_2$ are obtained from eqs.(\ref{F1},\ref{F2}) with $C_{ijkl}=0$. Eqs.(\ref{EinsteinF3},\ref{E4Fn}) apply to the case with flat space-time on the boundary. This is also the case studied in \cite{Safdi}. Recall that the author of \cite{Safdi} makes two further assumptions \cite{Safdi}. The first one is  $B_3=\frac{B_2-\frac{B_1}{2}}{3}$. And the second assumption is zero extrinsic curvature in the time-like direction. So we can drop the indices $(a,b,c,d)$ in eqs.(\ref{EinsteinF3},\ref{E4Fn}). We get
\begin{eqnarray}\label{Safdi16}
	S_{\Sigma}|_{\log}=\log(\ell/\delta)\, \int_{\Sigma} d^{4}x\sqrt{h_0}\,\left[\,
	2\,A\,
	E_{4}+3\pi B_1 (\frac{3}{2}T_1-T_2)-12\pi B_2(T_2)+6\pi B_3(T_3+9 T_1-12T_2)\,\right]，\nonumber\\
\end{eqnarray}
where the definitions of $T_n$ can be found in eq.(\ref{Ti}). Note that eq.(\ref{Safdi16}) reduces to the result of \cite{Safdi} eq.(\ref{6dlog}) when $B_3=\frac{B_2-\frac{B_1}{2}}{3}$. This is a non-trivial check of our results.

\subsection{ Logarithmic terms of EE from a general action}

In this sub-section, we investigate the universal terms of EE by using the general higher curvature gravity. We prove that it yields the same results as the above section. Our main method is the background-field approach developed in \cite{Miao1}. For simplicity, we focus on the action without the derivatives of the curvature $S(\hat{G}_{\mu\nu}, \hat{R}_{\mu\nu\sigma\rho})$. Besides, we assume this action has an asymptotically AdS solution.

We firstly expand the action around a referenced curvature $\bar{R}_{\mu\nu\rho\sigma}=-(\hat{G}_{\mu\rho}\hat{G}_{\nu\sigma}-\hat{G}_{\mu\sigma}\hat{G}_{\nu\rho})$.  
According to \cite{Miao1}, only the first few terms are relevant to the holographic Weyl anomaly and the logarithmic term of EE. We have
\begin{eqnarray}\label{mostgeneralaction}
	&&S(\hat{G}_{\mu\nu}, \hat{R}_{\mu\nu\sigma\rho})=\int
	d^{7}X\sqrt{-\hat{G}} [\ \sum_{n=0}^{3} \sum_{i=1}^{m_n} c^{(n)}_i \tilde{K}^n_i+O(\rho^4)\ ]\nonumber\\
	&&=\int
	d^{7}X\sqrt{-\hat{G}} [-\frac{c^{(1)}_1}{12}(\hat{R}+30)+c^{(2)}_1 \tilde{R}_{\mu\nu\rho\sigma}\tilde{R}^{\mu\nu\rho\sigma}+c^{(3)}_3\tilde{R}\tilde{R}_{\mu\nu\rho\sigma}\tilde{R}^{\mu\nu\rho\sigma}+c^{(3)}_6\tilde{R}_{\mu\nu}\tilde{R}^{\mu}_{\ \rho\sigma\beta}\tilde{R}^{\nu\rho\sigma\beta} \nonumber\\
	&&\ \ \ \ \ \ \ \ \ \ \ \ \ \ \ \ \ \ \ \  +c^{(3)}_7 M_2+c^{(3)}_8 M_1+O(\rho^4)\ ]\nonumber\\
\end{eqnarray}
where $c^{(n)}_i$ are some constants determined by the action and $m_n$ is the number of independent
scalars constructed from appropriate contractions of $n$ curvature tensors. For example, $m_1=1, m_2=3, m_3=8$. $\tilde{K}^n_i=K^n_i|_{[\hat{R}\rightarrow(\hat{R}-\bar{R})]}$ with
$K^n_i$ the independent scalars constructed from $n$
curvature tensors. For example, we have
\begin{eqnarray}\label{Kni}
	&&K^1_1=\hat{R},\nonumber\\
	&&K^2_i=(\hat{R}_{\mu\nu\rho\sigma}\hat{R}^{\mu\nu\rho\sigma},\
	\hat{R}_{\mu\nu}\hat{R}^{\mu\nu},\
	\hat{R}^2), \nonumber\\
	&&K^3_i=(\hat{R}^3,\hat{R}\hat{R}_{\mu\nu}\hat{R}^{\mu\nu},\hat{R}\hat{R}_{\mu\nu\rho\sigma}\hat{R}^{\mu\nu\rho\sigma},\hat{R}_{\mu}^{\nu}\hat{R}_{\nu}^{\rho}\hat{R}_{\rho}^{\mu},
	\hat{R}^{\mu\nu}\hat{R}^{\rho \sigma}\hat{R}_{\mu \rho \sigma\nu},
	\hat{R}_{\mu \nu}\hat{R}^{\mu \rho \sigma \lambda}\hat{R}^{\nu}_{\ \rho \sigma \lambda},\nonumber\\
	&&\ \ \ \ \ \ \ \ \hat{R}_{\mu \nu \rho \sigma}\hat{R}^{\mu \nu
		\lambda\chi}\hat{R}^{\rho \sigma}_{\ \ \lambda\chi},\hat{R}_{\mu \nu
		\rho\sigma}\hat{R}^{\mu\lambda\chi\sigma}\hat{R}^{\nu\ \ \rho}_{\
		\lambda\chi}),\nonumber
	\\&&...
\end{eqnarray}
For simplicity, we focus on the case with $c^2_1=0$ in this paper.  Without loss of generality, we set $c^1_1=-12, c^3_3=\lambda_3, c^3_6=\lambda_4, c^3_7=\lambda_2, c^3_8=\lambda_1$. Then the general action becomes
\begin{eqnarray}\label{generalaction}
	S=\int
	d^{7}X\sqrt{-\hat{G}}[\hat{R}+30+\lambda_1 M_1+\lambda_2 M_2+\lambda_3 \tilde{R}\tilde{R}_{\mu\nu\rho\sigma}\tilde{R}^{\mu\nu\rho\sigma}+\lambda_4 \tilde{R}_{\mu\nu}\tilde{R}^{\mu}_{\ \rho\sigma\beta}\tilde{R}^{\nu\rho\sigma\beta}+O(\rho^4)]\nonumber\\
\end{eqnarray}
Please refer to eq.(\ref{Mi}) and eq.(\ref{hatR}) for the defination of $M_n$ and $\tilde{R}$, respectively. According to \cite{Miao1}, the Weyl anomaly of dual CFTs is $\langle\,T^i{}_i\,\rangle=\sum_{n=1}^3 B_n\, I_n + 2 A \, E_6$ with central charges given by eq.(\ref{centralcharges})
\begin{eqnarray}\label{generalcentralcharges}
	&A&=\pi^3,\nonumber\\
	&B_1&=-\frac{1}{16}+\lambda_1,\nonumber\\
	&B_2&=-\frac{1}{64}+\lambda_2,\nonumber\\
	&B_3&=\frac{1}{192}.
\end{eqnarray}
Remarkably, the CFTs dual to the gravitational theories eq.(\ref{smartaction}) and eq.(\ref{generalaction}) have exactly the same central charges. This means that they must have the same universal terms of EE too. Thus $\tilde{R}\tilde{R}_{\mu\nu\rho\sigma}\tilde{R}^{\mu\nu\rho\sigma}$ and $\tilde{R}_{\mu\nu}\tilde{R}^{\mu}_{\ \rho\sigma\beta}\tilde{R}^{\nu\rho\sigma\beta}$ in the action eq.(\ref{generalaction}) can not contribute to universal terms of EE in order to be consistent with the results of sect. 4.1.1 and sect. 4.1.2.

Following the approach of sect. 4.1.1, we find that the Wald entropy of  $\tilde{R}\tilde{R}_{\mu\nu\rho\sigma}\tilde{R}^{\mu\nu\rho\sigma}$ and $\tilde{R}_{\mu\nu}\tilde{R}^{\mu}_{\ \rho\sigma\beta}\tilde{R}^{\nu\rho\sigma\beta}$ is indeed irrelevant to the universal terms of EE. However, mismatches come from the anomaly-like entropy if we choose $Q_{0z\bar{z}ij}$ and $T_0$ to be zero as in the original work of \cite{Dong}. This implies that the splittings of the conical metric eq.(\ref{ReviewTVQ}) are necessary.  Applying eqs.(\ref{HEEDong},\ref{splittingEinstein1},\ref{splittingEinstein2},\ref{6dlogCFTformula}-\ref{6dlogCFTformula3}), we get the anomaly-like entropy as
\begin{eqnarray}\label{AnomalyRR}
	S_{\text{Anomaly}}&=&-\int d\rho d^4y \sqrt{h}\big{[}\lambda_3 K_{zij}K_{\bar{z}}^{\ ij}(3K_{amn}K^{amn}-K^aK_a-2Q_{0a}^{a}{}^m_m+24T_0)\nonumber\\
	&+& \lambda_4K_{zij}K_{\bar{z}}^{\ ij}(K_{zmn}K_{\bar{z}}^{\ mn}-Q_{0\ z\bar{z}m}^{\ \ \ \ \ m}+6T_0) +\frac{\lambda_4}{2}K_{zil}K_{\bar{z}\ j}^{\ l}(2K^{aim}K_{am}^{\ \ j}-K_aK^{aij}-Q_a^{aij})\big{]} +...\nonumber\\
	&=&0 \ \log(l/\delta)+...
\end{eqnarray}
where `...' denotes terms irrelevant to the logarithmic terms of EE. In the above derivations, we have used the splittings eqs.(\ref{splittingEinstein1},\ref{splittingEinstein2}) and the fact that only the $tr K^4$ and $(tr K^2)^2$ of the $O(K^4)$ terms contribute to the universal term of EE for 6d CFTs. Now it is clear that $\tilde{R}\tilde{R}_{\mu\nu\rho\sigma}\tilde{R}^{\mu\nu\rho\sigma}$ and $\tilde{R}_{\mu\nu}\tilde{R}^{\mu}_{\ \rho\sigma\beta}\tilde{R}^{\nu\rho\sigma\beta}$ indeed do not contribute to the logarithmic terms. So the higher curvature gravity with $c^2_1=0$ gives the same universal terms of EE as those of sect. 4.1.1 and sect. 4.1.2.  As for the case with $c^2_1$ non-zero, the calculation is quite complicated. But there is no indication that this case would give a different result. We leave the check of this case as an exercise for the readers. Finally, it should be mentioned that, in addition to equations of motion, eq.(\ref{AnomalyRR}) can be regarded as another derivation of the splittings eqs.(\ref{splittingEinstein1},\ref{splittingEinstein2}). That is because different higher curvature gravity must give the same formula of universal terms of EE. Therefore, the logarithmic terms of  eq.(\ref{AnomalyRR}) must be zero.

\section{Field theoretical method}

In this section, we compute the universal terms of EE by using the field theoretical method and then compare with the holographic results.  Similar to the bulk case, we meet the splitting problem. Since now we do not know how to fix the splitting problem on the boundary, we assume the most general expressions.
We find that there indeed exists suitable splittings which could make the holographic and the field theoretical results match. 

Recall that Weyl anomaly for 6d CFTs is given by
\be
\langle\,T^i{}_i\,\rangle=\sum_{n=1}^3 B_n\, I_n + 2 A \, E_6,
\label{Fieldtrace6}
\ee
where $E_6$ is the Euler density and $I_i$ are conformal invariants defined by
\bea
I_1&=&C_{k i j l}C^{i m n j}C_{m~~\,n}^{~~k l} ~, \qquad
I_2=C_{i j}^{~~k l}C_{k l}^{~~m n}C^{~~~i j}_{m n}~,\label{Fieldtrace6x1} \\
I_3&=&C_{i k l m}(\nabla^2 \, \de^{i}_{j}+4R^{i}{}_{j}-{6 \over 5}\,R
\,
\de^{i}_{j})C^{j k l m}\,.\label{Fieldtrace6x2}
\eea
In the field theoretical approach, one can derive the universal terms of EE from the Weyl anomaly. Take the Weyl anomaly as a gravitational action and then calculate the `entropy' of this `action'. It turns out that this 'entropy'  equals to the logarithmic term of EE for CFTs \cite{Solodukhin,Hung}.

\subsection{$F_1$ and $F_2$}

Let us firstly study the case of $F_1$ and $F_2$. We find that the field theoretical results exactly match the holographic ones for the $C^2$ and $C k^2$ terms. As for the $k^4$ terms, one meet with the splitting problem for $q_{0z\bar{z}ij}$ and $t_0$. 
Since now we do not know how to fix the splitting for $t, q$ on the boundary, we assume the following general expressions
\begin{eqnarray}\label{q0t0}
	&&t_0=z_1 k_{a mn}k^{amn}+z_2 k_a k^a\nonumber\\
	&&q_{0\ z\bar{z}ij}=(x_1 k_{zim}k^{\ m}_{\bar{z}\ \ j}+x_2\  g_{ij} k_{z mn}k_{\bar{z}}^{\ mn}+y_1 k_z k_{\bar{z}ij}+y_2\  g_{ij}k_zk_{\bar{z}}) +c.c.
\end{eqnarray} 
Recall that, in sect.4.1.1, we have already proved that the field theoretical results match the holographic ones for Wald entropy ($C^2$ terms), so we focus on the anomaly-like entropy below. 

Applying the formula eq.(\ref{HEEDong}), we get the anomaly-like entropy for $I_1$ eq.(\ref{Fieldtrace6x1}) as
\begin{eqnarray}\label{AnomalyofI1}
	&&S_{1}=\int_{\Sigma}d^4y \sqrt{h_0} \big{[}\ 24\pi  \bar{k}_{zij}\bar{k}_{\bar{z}mn} C^{imjn}-12\pi  \bar{k}_{zij}\bar{k}_{\bar{z}mn} C_0^{imjn}\nonumber\\
	&&\ \ \ \ \ \ \ \ \ \ -96\pi \bar{k}_{zi}^m\bar{k}_{\bar{z}mj} C^{ij}_{\ \ z\bar{z}}+48\pi \bar{k}_{zi}^m\bar{k}_{\bar{z}mj} C_0^{ij}{}_{z\bar{z}}\nonumber\\
	&&\ \ \ \ \ \ \ \ \ \  96\pi \bar{k}_{zmn}\bar{k}_{\bar{z}}^{\ mn}C_{z\bar{z}z\bar{z}}-48\pi \bar{k}_{zmn}\bar{k}_{\bar{z}}^{\ mn}C_{0\ z\bar{z}z\bar{z}}\  \big{]}
\end{eqnarray}
where $\bar{k}_{aij}$ is the traceless part of the extrinsic curvature and $C_0\sim k^2$ is defined in the Appendix. C. Comparing eq.(\ref{AnomalyofI1}) with eq.(\ref{FA1}), we find that the $Ck^2$ terms match exactly. If we require that the $k^4$ terms also match, we get a unique solution to eq.(\ref{q0t0}) 
\begin{eqnarray}\label{xsolutionCFT}
	x_1=1,\  x_2=\frac{1}{4}-6 z_1,\ y_1=0,\ y_2=-\frac{1}{16} -6 z_2
\end{eqnarray}

Let us go on to compute the anomaly-like entropy for $I_2$ eq.(\ref{Fieldtrace6x1}). Using eq.(\ref{HEEDong}), we obtain
\begin{eqnarray}\label{AnomalyofI2}
	S_{2}&=&\int_{\Sigma}d^4y \sqrt{h_0} \big{[}\ -384 \pi \bar{k}_{z\ m}^{\ i}\bar{k}_{\bar{z}}^{\ jm} C_{zj\bar{z}i}+192\pi \bar{k}_{z\ m}^{\ i}\bar{k}_{\bar{z}}^{\ jm} C_{0\ zj\bar{z}i}\ \big{]},
\end{eqnarray}
where $C_0\sim k^2$ is given by eq.(\ref{Weyltensor1}). Similar to the case of $I_1$, the $Ck^2$ terms of eq.(\ref{FA2}) and eq.(\ref{AnomalyofI2}) match exactly. The $k^4$ terms also match if we impose the condition eq.(\ref{xsolutionCFT}). This is a non-trivial self-consistent testing of the splittings eq.(\ref{xsolutionCFT}) on the boundary. Note that comparing the holographic results and the field theoretical results for $F_1$ and $F_2$ does not fix $z_1, z_2$. 

To end this section, we show some details of the derivation of eq.(\ref{xsolutionCFT}). For simplicity, we focus on the case of vanishing extrinsic curvature in the time-like direction (one can check that the general case gives the same results). Then we can replace $k_{aij}$ by $\frac{1}{2}k_{ij}$. From eqs.(\ref{AnomalyofI1},\ref{AnomalyofI2},\ref{WeyltensorSafdi}), we can derive the $k^4$ terms as
\begin{eqnarray}\label{CFTlogtry1}
	&&B_1S_1+B_2S_2\nonumber\\
	&=&\int_{\Sigma}d^4y\sqrt{h_0} \big{[}\ 3\pi[B_1(x_1-2)-4B_2 x_1] tr k^4-\frac{3\pi}{2}[B_1(x_1-2y_1-3)-4B_2(1+x_1-2y_2)] k tr k^3\nonumber\\
	&&+\frac{3}{20}\pi[B_1(21+2x_1+28x_2+168z_1)+4B_2(1+2x_1-12x_2-72z_1)] (tr k^2)^2\nonumber\\
	&&+\frac{3}{160}\pi[B_1(19+6y_1-56y_2-336z_2)+4B_2(9-14y_1+24y_2+144z_2)] k^4\nonumber\\
	&&+\frac{3}{80}\pi[B_1(-79+3x_1-28x_2-32y_1+112y_2-168z_1+672z_2)\nonumber\\
	&&\ \ \ \ \ \ \ \ \ -4B_2(29+7x_1-12x_2-48y_1+48y_2-72z_1+288z_2)] k^2tr k^2\ \big{]}
\end{eqnarray}
For 6d CFTs with $B_3=0$, the holographic $k^4$ terms eq.(\ref{Safdi16}) becomes
\begin{eqnarray}\label{6dlogflat}
	S_{\Sigma}|_{\log}&=&\log(\ell/\delta)\, \int_{\Sigma} d^{4}y\sqrt{h_0}\,\left[\,
	3\pi B_1 (\frac{3}{2}T_1-T_2)-12\pi B_2(T_2)\right]\nonumber\\
	&=&\log(\ell/\delta) \int_{\Sigma}d^{4}y\sqrt{h_0} 3\pi \big{[} -(B_1+4B_2) tr k^4+(B_1+4B_2)k tr k^3+\frac{3}{2}B_1 (tr k^2)^2\nonumber\\
	&&\ \ \ \ \ \ \ \ \ \ \ \ \ \ \ \ \ \ \ \ \ \ \ \ \ \ \ \ \ \ \ -\frac{3}{8}(3B_1+4B_2) k^2tr k^2 +\frac{3}{64}(3B_1+4B_2) k^4 \big{]}.
\end{eqnarray}
Compare eq.(\ref{CFTlogtry1}) with eq.(\ref{6dlogflat}), we find a unique solution
\begin{eqnarray}
	x_1=1,\  x_2=\frac{1}{4}-6 z_1,\ y_1=0,\ y_2=-\frac{1}{16} -6 z_2
\end{eqnarray} 
Note that $B_1$ and $B_2$ are independent central charges, so there are ten equations (\ref{CFTlogtry1}) for six unkown parameters. Thus it is non-trivial that we have consistent solutions. 

\subsection{$F_3$}

Now let us go on to study the $F_3$ term. In sect. 4.1.2, we have discussed the holographic $F_3$ for two interesting cases. In this first case we set $k_{aij}=0$ and derive the $C^2$  terms of $F_3$ eq.(\ref{SEinstein1}). And in the second case, we focus on the flat boundary spacetime and obtain the $k^4$ terms of $F_3$ eqs.(\ref{SEinstein1},\ref{Safdi16}). In this section, we calculate the corresponding field theoretical results and compare with the holographic ones. We find that the $C^2$ terms of $F_3$ indeed match. This can be regarded as a resolution of the HMS puzzle \cite{Hung, Miao2}. While for the $k^4$ terms, we have to deal with the splitting problem. We assume eqs.(\ref{q0t0},\ref{xsolutionCFT}) and check if this assumption of splitting could pass the $F_3$ test or not.

Case {\bf I}:  $k_{aij}=0$

Let us firstly investigate the case with zero extrinsic curvature. It is found by HMS \cite{Hung} that there are mismatches between the holographic and the field theoretical universal terms of EE even for the entangling surfaces with zero extrinsic curvature. Recently, the authors of \cite{Miao2} find that HMS have ignored the anomaly-like entropy from the Weyl anomaly $I_3$. After taking into account this contribution, the holographic and CFT universal terms of EE indeed match \cite{Miao2}. For simplicity \cite{Hung,Miao2} both focus on the cases with $k_{aij}=0$ and $R_{abci}=3\epsilon_{ab}V_{ci}=0$. Here we drop the second constraint $R_{abci}=3\epsilon_{ab}V_{ci}=0$ and check if the holographic and the field theoretical results still match. We only need to compare $\Delta S$ eq.(\ref{EinsteinDeltaS}) with the anomaly-like entropy from $I_3$. That is because the anomaly-like entropy of $I_1$ and $I_2$ vanishes for $k_{aij}=0$. Note further that the anomaly-like entropy of $I_3$ only comes form $C_{ijkl}\Box C^{ijkl}\cong-\nabla_{m} C_{ijkl}\nabla^{m}C^{ijkl}$ for the case of zero extrinsic curvature. 

When the extrinsic curvature vanishes, the splitting problem disappears and the anomaly-like entropy for the gravitational action with one derivative of the curvature is given by \cite{Miao2}
\begin{eqnarray}\label{HMSHEEsix}
	S_{Anomaly}&=&2\pi\int d^(D-2)y \sqrt{h}\big{[}64\big{(}\frac{\partial^2L}{\partial \nabla_zR_{zizl}\partial \nabla_{\bar{z}}R_{\bar{z}k\bar{z}l}}\big{)}_{\alpha_1}\frac{Q_{zzij}Q_{\bar{z}\bar{z}kl}}{\beta_{\alpha_1}}\nonumber\\
	&+&96 i\big{(}\frac{\partial^2L}{\partial \nabla_zR_{zizl}\partial \nabla_{\bar{z}}R_{\bar{z}z\bar{z}k}}\big{)}_{\alpha_1}\frac{Q_{zzij}V_{\bar{z}k}}{\beta_{\alpha_1}}+c.c\nonumber\\
	&+&144\big{(}\frac{\partial^2L}{\partial \nabla_zR_{z\bar{z}zl}\partial \nabla_{\bar{z}}R_{\bar{z}z\bar{z}k}}\big{)}_{\alpha_1}\frac{V_{zl}V_{\bar{z}k}}{\beta_{\alpha_1}}
	\big{]},
\end{eqnarray}
where $Q, V$ are defined in the conical metric
\begin{eqnarray}\label{HMScone}
	ds^2=e^{2A}[dzd\bar{z}+e^{2A}T(\bar{z}dz-zd\bar{z})^2]+2i e^{2A} V_i(\bar{z}dz-zd\bar{z})dy^i\nonumber\\
	+( h_{ij}+Q_{ij})dy^idy^j.
\end{eqnarray}
Here $A=-\frac{\epsilon}{2}\lg(z\bar{z}+b^2)$ is regularized warp factor and $V_i, Q_{ij}$ are defined as
\begin{eqnarray}\label{HMSTVQ}
	&&V_i=U_i+z V_{zi}+\bar{z}V_{\bar{z}i}+O(z^2),\nonumber\\
	&&Q_{ij}=z^2Q_{zzij}+\bar{z}^2Q_{\bar{z}\bar{z}ij}+2z \bar{z} e^{2A} Q_{z\bar{z}ij}+O(z^3)
\end{eqnarray}
Applying the formula eq.(\ref{HMSHEEsix}), we derive the anomaly-like entropy of  $C_{ijkl}\Box C^{ijkl}\cong-\nabla_{m} C_{ijkl}\nabla^{m}C^{ijkl}$ as
\begin{eqnarray}\label{HMSCBoxC}
	S_{A}&=&\int d^4y \sqrt{h_0}\big{[}128\pi \bar{Q}_{zzij}\bar{Q}_{\bar{z}\bar{z}}^{\ \ ij}+432\pi V_{zi}V_{\bar{z}}^{\ i}\big{]}.
\end{eqnarray}
It should be mentioned that the total entropy of $\Box C_{ijkl}C^{ijkl}$ vanishes by using the approach of \cite{Dong, Miao2}.

Substituting the conical metric eq.(\ref{HMScone}) with $A=0$ into $\Delta S$ eq.(\ref{EinsteinDeltaS}), we get
\begin{eqnarray}\label{HMDeltaS}
	\Delta S&=&\big{[}128\pi \bar{Q}_{zzij}\bar{Q}_{\bar{z}\bar{z}}^{\ \ ij}+432\pi V_{zi}V_{\bar{z}}^{\ i}\big{]}.
\end{eqnarray}
which is exactly the same as eq.(\ref{HMSCBoxC}). Thus the holographic and the field theoretical  results match for the $C^2$ terms of $F_3$.

Case {\bf II}:  flat $\overset{\scriptscriptstyle{(0)}}{g_{ij}}$

Now let us go on to study the case with flat spacetime on the boundary. The holographic result of $2\pi F_3$ is given by eq.(\ref{Safdi16})
\begin{eqnarray}\label{F3flatcompare}
	S_{\Sigma}|_{\log}=\log(\ell/\delta)\, \int_{\Sigma} d^{4}y\sqrt{h_0}\,\left[\,
	6\pi (T_3+9 T_1-12T_2)\,\right]，
\end{eqnarray}
with 
\begin{eqnarray}
	T_1=(\text{tr} \bar{k}^2)^2,\ T_2=\text{tr} \bar{k}^4, \ T_3=(\nabla_a k)^2-\frac{25}{16}k^4+11k^2 tr k^2-6(tr k^2)^2-16 k tr k^3+12 tr k^4.\nonumber\\
\end{eqnarray}

Applying the method developed in \cite{Dong,Miao2} together with the splittings eqs.(\ref{q0t0},\ref{xsolutionCFT}), we can derive $2\pi F_3$ as the entropy of $I_3$.  We list the results below.

{\bf  I} For $ds^2=dzd\bar{z}+(1+\frac{z+\bar{z}}{2})^2dy_1^2+dy_2^2+dy_3^2+dy_4^2$,
we obtain the entropy of $I_3$ as
\begin{eqnarray}\label{F31}
	S_{I}|_{\log}= \int_{\Sigma} d^{4}y\sqrt{h_0}\  \frac{27 \pi }{8}
\end{eqnarray}
which agrees with the holographic result eq.(\ref{F3flatcompare}) with $k^{\hat{i}}_{\ \hat{j}}=\text{diag}\{1,0,0,0 \}$. 

{\bf  II} For $ds^2=dzd\bar{z}+(1+\frac{z+\bar{z}}{2})^2(dy_1^2+\sin^2y_1dy_2^2)+dy_3^2+dy_4^2$, we derive the entropy of $I_3$ as
\begin{eqnarray}\label{F32}
	S_{II}|_{\log}=\int_{\Sigma} d^{4}y\sqrt{h_0}\  30\pi,
\end{eqnarray}
which matches the holographic result eq.(\ref{F3flatcompare}) with $k^{\hat{i}}_{\ \hat{j}}=\text{diag}\{1,1,0,0 \}$.

{\bf  III} For $ds^2=dzd\bar{z}+(1+\frac{z+\bar{z}}{2})^2(dy_1^2+\sin^2y_1dy_2^2+\sin^2y_1\sin^2y_2dy_3^2)+dy_4^2$, we get the entropy of $I_3$ as
\begin{eqnarray}\label{F33}
	S_{III}|_{\log}=\int_{\Sigma} d^{4}y\sqrt{h_0}\  \frac{459 \pi }{8},
\end{eqnarray}
which is consistent with the holographic result eq.(\ref{F3flatcompare}) with $k^{\hat{i}}_{\ \hat{j}}=\text{diag}\{1,1,1,0 \}$.

{\bf IV} For  $ds^2=dzd\bar{z}+(1+\frac{z+\bar{z}}{2})^2(dy_1^2+\sin^2y_1dy_2^2+\sin^2y_1\sin^2y_2dy_3^2+\sin^2y_1\sin^2y_2\sin^2y_3^2dy_4^2)$, we have the entropy of $I_3$
\begin{eqnarray}\label{F34}
	S_{IV}|_{\log}=0,
\end{eqnarray}
which also agrees with the holographic result eq.(\ref{F3flatcompare}) with $k^{\hat{i}}_{\ \hat{j}}=\text{diag}\{1,1,1,1 \}$.  

Now it is clear that the splittings eq.(\ref{q0t0},\ref{xsolutionCFT}) have passed the $F_3$ test.  Remarkably, we cannot fix the splittings completely by comparing the holographic and field theoretical universal terms of EE. It seems that we have more than one way to split the conical metrics on the boundary and such freedom does not affect the universal terms of EE.

\section{Conclusions}

We have investigated the universal terms of EE for 6d CFTs by applying holographic and the field theoretical methods, respectively. Our results agree with those of \cite{Hung, Safdi}. We find the holographic and the field theoretical results match for the $C^2$ and $Ck^2$ terms. While for the $k^4$ terms, we meet the splitting problem for the conical metrics. We fix the splitting problem in the bulk by using two different methods. The first one is by using equations of motion and second one is requiring that all the higher derivative theories of gravity yield the same logarithmic terms of EE. These two methods give consistent results for the splitting in the bulk. As for the splitting on the boundary, we assume the general forms and find there indeed exists suitable splitting which can make the holographic and CFT $k^4$ terms match. Since we have much more equations than the free parameters, this match is non-trivial. Remarkably, we can not fix the splitting on the boundary completely by comparing the holographic and field theoretical results. It seems that we have some freedom to split the conical metrics on the boundary and such freedom does not affect the universal terms of EE for CFTs. That is not surprising, since the terms (Weyl anmoly) we studied are quite special. Actually, for Lovelock gravity, arbitrary splitting would not affect the entropy. How to fix the splitting problem on the boundary is an interesting problem. For the cases with gravity duals, we could obtain the conical metrics on the boundary from the one in the bulk. While for the general cases, now it is not clear to us how to fix this problem. We hope to address this problem in future. Finally, we want to point out how much our holographic results $F_i$ eqs.(\ref{F1},\ref{F2},\ref{GeneralF3}) depend on the splittings. It turns out that that the combinations $(F_3-3F_2-12F_1)$ and $(2F_1+F_2)$ are independent of the splittings, due to the fact that they can be derived from the holographic entanglement entropy of Einstein gravity and Lovelock gravity which are irrelevant to the splittings. In other words, our results do not depend on the splittings when the central charges satisfy $B_3=\frac{B_2-\frac{B_1}{2}}{3}$. Without loss of generality, we choose $F_2$ as the third independent combination of $F_i$. As mentioned above, the splitting problem does not affect the $C^2$ and $C k^2$ terms. Thus, only the $k^4$ terms of $F_2$ are relevant to the splitting problem.

\section*{Acknowledgements}

R. X. Miao is supported by Sino-German (CSC-DAAD) Postdoc Scholarship Program. R. X. Miao thank S. Theisen and X. Dong for helpful discussions, and also J. Camps for pointing out a typo. 

\appendix

\section{Universal relations from extremal entropy condition}

In this section, we derive the universal identity eq.(\ref{rules1}) by taking the variation of the entropy functional. For simplicty, we focus on Einstein gravity in asymptotically AdS space-time. The basic idea is to study the variation of the universal logarithmic terms of the entropy functional. 

Recall that the embedding functions of the entangling surface $m$ into the bulk are given by
\begin{eqnarray}\label{Appembeddingfunctions}
	X^i(\tau,y^{\hat{j}})=\overset{\scriptscriptstyle{(0)}}{X^i}(y^{\hat{j}})+\overset{\scriptscriptstyle{(1)}}{X^i}(y^{\hat{j}})\tau+\overset{\scriptscriptstyle{(2)}}{X^i}(y^{\hat{j}})\tau^2+...
\end{eqnarray}
from which we can derive the induced metric on $m$ as
\begin{eqnarray}\label{Appinducedhab1}
	h_{\tau\tau}&=&\frac{1}{4\tau^2}\Big{(}1+...+4(d-2)\,
	\overset{\scriptscriptstyle{(\frac{d-2}{2})}}{X^i}\overset{\scriptscriptstyle{(1)}}{X^i}\overset{\scriptscriptstyle{(0)}}{g}_{ij}\,\tau^{\frac{d-2}{2}}+\cdots\Big{)},\\
	h_{\hat{i}\hat{j}}&=&\frac{1}{ \tau}\Big{(}\overset{\scriptscriptstyle{(0)}}{h}_{\hat{i}\hat{j}}+...+(2\partial_{(\hat{i}}\overset{\scriptscriptstyle{(\frac{d-2}{2})}}{X^m}\partial_{\hat{j})}\overset{\scriptscriptstyle{(0)}}{X^n}\overset{\scriptscriptstyle{(0)}}{g_{mn}}+\partial_{\hat{i}}\overset{\scriptscriptstyle{(0)}}{X^m}\partial_{\hat{j}}\overset{\scriptscriptstyle{(0)}}{X^n}\overset{\scriptscriptstyle{(\frac{d-2}{2})}}{X^k}\partial_k \overset{\scriptscriptstyle{(0)}}{g_{mn}})\,\tau^{\frac{d-2}{2}}+...\Big{)},\label{Appinducedhab2}
\end{eqnarray}
where we only list terms including $\overset{\scriptscriptstyle{(\frac{d-2}{2})}}{X^i}$ in the above equations. 
Remarkably, only the lineared terms of $\overset{\scriptscriptstyle{(\frac{d-2}{2})}}{X^i}$ appear in the  logarithmic terms of the  entropy functional $\frac{1}{4G}\int_{\delta^2}d\tau\int_{\Sigma}d^{d-2}y\sqrt{h}$

\begin{eqnarray}\label{APPX2Einstein}
	S_{\text{log}}
	&=&\frac{1}{4G}\log(\ell/\delta)\int_{\Sigma} d^{d-2}y \sqrt{h_0}[2(d-2)\overset{\scriptscriptstyle{(\frac{d-2}{2})}}{X^i} \overset{\scriptscriptstyle{(1)}}{X^j} \overset{\scriptscriptstyle{(0)}}{g_{ij}}+\overset{\scriptscriptstyle{(0)}}{h^{\hat{i}\hat{j}}}\partial_{\hat{i}}\overset{\scriptscriptstyle{(0)}}{X^m}\partial_{\hat{j}}\overset{\scriptscriptstyle{(\frac{d-2}{2})}}{X^n}\overset{\scriptscriptstyle{(0)}}{g_{mn}}+\frac{1}{2}\overset{\scriptscriptstyle{(0)}}{h^{ij}}\overset{\scriptscriptstyle{(\frac{d-2}{2})}}{X^k}\partial_k \overset{\scriptscriptstyle{(0)}}{g_{ij}}]+...\nonumber\\
\end{eqnarray}
where $'...'$ denote terms without $\overset{\scriptscriptstyle{(\frac{d-2}{2})}}{X^i}$. Taking the variation of $ \overset{\scriptscriptstyle{(\frac{d-2}{2})}}{X^i}$ for eq.(\ref{APPX2Einstein}), we get
\begin{eqnarray}\label{Appkimn} \overset{\scriptscriptstyle{(1)}}{X^j}&=&\frac{1}{2(d-2)}\overset{\scriptscriptstyle{(0)}}{h}^{\hat{m}\hat{n}}(\partial_{\hat{m}}\partial_{\hat{n}}\overset{\scriptscriptstyle{(0)}}{X^j}-\overset{\scriptscriptstyle{(0)}}{\gamma^{\hat{l}}}_{\hat{m}\hat{n}}\partial_{\hat{l}}\overset{\scriptscriptstyle{(0)}}{X^j}+\overset{\scriptscriptstyle{(0)}}{\Gamma^j}_{kl}\partial_{\hat{m}}\overset{\scriptscriptstyle{(0)}}{X^k}\partial_{\hat{n}}\overset{\scriptscriptstyle{(0)}}{X^l})\nonumber\\
	&=&\frac{1}{2(d-2)}k^j
\end{eqnarray}
where $\gamma^{\hat{l}}_{\hat{m}\hat{n}}$ and $\Gamma^{j}_{kl}$ are the Levi-Civita connection with respect to $\overset{\scriptscriptstyle{(0)}}{h}_{\hat{i}\hat{j}}$ and $\overset{\scriptscriptstyle{(0)}}{g}_{ij}$, respectively.

Now we finish the derivations of eq.(\ref{rules1}) from the extremal entropy condition. Although we only studied the case of Einstein gravity , similar to $\overset{\scriptscriptstyle{(1)}}{g}_{ij}$ eq.(\ref{g1}) it is expected that our approaches and conclusions of this section can be generalized to the general higher derivative gravity, due to the fact that the universal relation eq.(\ref{rules1}) can be derived from PBH transformation \cite{Theisen2}.  This means that, at the leading order, the asymptotic symmetry forces that the extremal entropy surface (bulk entangling surface ) approaches the minimal area surface near the AdS boundary. Of course, they can be different at the subleading orders near the AdS boundary, since $\overset{\scriptscriptstyle{(n)}}{X}{}^i$ with $n\ge 2$ are non-universal. Fortunately, we do not need $\overset{\scriptscriptstyle{(n)}}{X}{}^i$ with $n\ge 2$ for the derivations of universal terms of EE for 4d and 6d CFTs, similar to the case that we do not need $\overset{\scriptscriptstyle{(n)}}{g}_{ij}$ with $n\ge 2$ for the calculations of holographic Weyl anomaly for 4d and 6d CFTs \cite{Miao1}.

\section{The conformal invariance of $F_3$}

In this section, we prove that the logarithmic terms of EE for Einstein gravity $S_E$ eq.(\ref{SEinstein}) are conformally invariant. Recall that $F_3$ is a combination of $S_E$ and the conformal invariants $F_1, F_2, E_4$. Thus, equivalently, we shall prove $F_3$ is conformally invariant. For simplicity, we focus on the infinitesimal conformal transformations. According to \cite{Theisen}, we have 
\begin{eqnarray}\label{Con1}
	&&\delta \overset{\scriptscriptstyle{(0)}}{g}_{ij}=2\sigma \overset{\scriptscriptstyle{(0)}}{g}_{ij}\nonumber\\
	&&\delta \overset{\scriptscriptstyle{(1)}}{g}_{ij}=\nabla_i\nabla_j\sigma\nonumber\\
	&&\delta \overset{\scriptscriptstyle{(2)}}{g}_{ij}=-2\sigma \overset{\scriptscriptstyle{(2)}}{g}_{ij}+\frac{1}{2}\nabla^k\sigma\nabla_k \overset{\scriptscriptstyle{(1)}}{g}_{ij}-\frac{1}{2}\nabla^m\sigma\nabla_{(i}\overset{\scriptscriptstyle{(1)}}{g}_{j)m}+\frac{1}{2}\overset{\scriptscriptstyle{(1)}}{g}_{(i}^{\ k}\nabla_{j)}\nabla_k\sigma.\nonumber\\
\end{eqnarray}
and 
\begin{eqnarray}\label{Con2}
	&&\delta k^m_{\ ij}=-h_{ij}\tilde{g}^{\perp}{}^{m n}\nabla_n \sigma\nonumber\\
	&&\delta k^m=-2\sigma k^m-4\tilde{g}^{\perp}{}^{m n}\nabla_n \sigma\nonumber\\
	&&\delta \Gamma^m_{\ ij}=\delta^m_i \nabla_j \sigma+\delta^m_j \nabla_i \sigma- \overset{\scriptscriptstyle{(0)}}{g}_{ij} \nabla^m \sigma\nonumber\\
	&&\delta R_{ijkl}=2\sigma R_{ijkl}+\overset{\scriptscriptstyle{(0)}}{g}_{il} \nabla_j \nabla_k\sigma-\overset{\scriptscriptstyle{(0)}}{g}_{ik} \nabla_j \nabla_l\sigma+\overset{\scriptscriptstyle{(0)}}{g}_{jk} \nabla_i \nabla_l\sigma-\overset{\scriptscriptstyle{(0)}}{g}_{jl} \nabla_i \nabla_k\sigma
\end{eqnarray}

Substituting eqs.(\ref{Con1},\ref{Con2}) into eq.(\ref{SEinstein}), we get
\begin{eqnarray}\label{deltaSEinstein}
	\delta_{\sigma}  S_{E}&=&\pi\log(\ell/\delta)\int_{\Sigma} d^4y \sqrt{h_0}[ \overset{\scriptscriptstyle{(1)}}{g}^{ij}h_{ij}k^m\nabla_m \sigma+\frac{1}{4}h^{ij}k^m\nabla_m\nabla_i\nabla_j\sigma+\frac{1}{4}\tilde{g}^{\perp}{}^{m n}h^{ij}k^lR_{limj}\nabla_n\sigma\nonumber\\
	&&-2\overset{\scriptscriptstyle{(1)}}{g}^{ij}k^m_{\ ij}\nabla_m \sigma+h^{ij}\nabla^m\overset{\scriptscriptstyle{(1)}}{g}_{ij}\nabla_m\sigma-\tilde{g}^{\perp}{}^{m n}h^{ij}\nabla_m\sigma \nabla_n \overset{\scriptscriptstyle{(1)}}{g}_{ij}-\frac{3}{32}k^mk_mk^n\nabla_n \sigma\nonumber\\
	&&+\frac{1}{4}k^mk^n\nabla_m\nabla_n\sigma-\overset{\scriptscriptstyle{(1)}}{g}_{ij}k_i\nabla_j\sigma+\frac{1}{2}k^mk_{mij}k^{n ij}\nabla_n\sigma-\overset{\scriptscriptstyle{(1)}}{g}^i_mh^{mj}\nabla_i\nabla_j\sigma+\overset{\scriptscriptstyle{(1)}}{g}^{ij}h_{ij}h^{mn}\nabla_m\nabla_n\sigma\nonumber\\
	&&-h^{ij}\nabla^k\sigma\nabla_i\overset{\scriptscriptstyle{(1)}}{g}_{jk}+\frac{1}{16}h^{ij}k^m\nabla_ik_j \nabla_m\sigma-\frac{1}{16}h^{ij}k^m\nabla_ik_m \nabla_j\sigma+\frac{1}{2}h^{ij}\nabla_ik^m\nabla_j\nabla_m\sigma\nonumber\\
	&&-\frac{5}{32}k^mk_mh^{ij}\nabla_i\nabla_j\sigma+\frac{1}{2}k_mk^{mij}\nabla_i\nabla_j\sigma-2\overset{\scriptscriptstyle{(1)}}{g}^{ij}h_{ik}\nabla^k \tilde{g}^{\perp}{}_{jl}\nabla^l\sigma-\frac{1}{4}h^{ij}\nabla_ik^m\nabla_j\tilde{g}^{\perp}{}_{m n}\nabla_n\sigma\nonumber\\
	&&-\frac{1}{4}\tilde{g}^{\perp}{}^{ij}h^{kl}\nabla_kk_i\nabla_j\nabla_l\sigma].
\end{eqnarray}
Let us try to simplify the above complicated results. The trick is to replace the covariant derivative $\nabla_i$ with respect to $\overset{\scriptscriptstyle{(0)}}{g}_{ij}$ by the intrinsic covariant derivative $D_i$ with respect to $h_{ij}$ as much as possible. Besides, we find the following formulas are useful:
\begin{eqnarray}\label{trick}
	&&h_i^mh_j^n\nabla_m V_n=D_i (h_j^nV_n)-k^m_{\ ij}V_m\nonumber\\
	&&h_i^mh_j^n\nabla_m \nabla_n\sigma=D_i D_j\sigma-k^m_{\ ij}\nabla_m\sigma\nonumber\\
	&&h_n^m\nabla_m h_{ij}=k_{inj}+k_{jni}\nonumber\\
	&&k_m\nabla_n h^{mn}=k^mk_m\nonumber\\
	&&k^mh_i^ph_j^qh_k^lR_{mpql}=k^m(\nabla_{j}k_{mik}-\nabla_{k}k_{mij})\nonumber\\
	&&k^mh^{ik}h^{jl}R_{mijk}\nabla_l\sigma=\frac{1}{2}D_i\sigma D^i (k^mk_m)-D^i\sigma D^j(k^mk_{mij})+\nabla_ik_mk^{mij}\nabla_j\sigma\nonumber\\
	&&h^{ij}k^m\nabla_i\nabla_m\nabla_j\sigma=D^i(h_i^jk^m\nabla_m\nabla_j\sigma)-k^mk^n \nabla_i\nabla_j\sigma-h^{ij}\nabla_ik^m\nabla_m\nabla_j\sigma
\end{eqnarray}
Applying the above formulas, we can simplify $\delta_{\sigma}  S_E$ as
\begin{eqnarray}\label{deltaSEinstein1}
	\delta_{\sigma} S_{E}&=&\pi\log(\ell/\delta)\int_{\Sigma} d^4y \sqrt{h_0}D^i [\frac{1}{4}h_i^jk^m\nabla_m\nabla_j\sigma+ \overset{\scriptscriptstyle{(1)}}{g}_{mn}h^{mn}D_i\sigma -h_{ij}\overset{\scriptscriptstyle{(1)}}{g}^{jm}\nabla_m\sigma\nonumber\\
	&&-\frac{1}{8}k^mk_mD_i\sigma+\frac{1}{4}k^mk_{mij}D^j\sigma].
\end{eqnarray}
which are just total derivative terms. Now it is clear that $S_E$ eq(\ref{SEinstein}) and thus $F_3$ eq.(\ref{GeneralF3}) are conformally invariant up to some total derivative terms. 

\section{Weyl tensor}

The Weyl tensor in D-dimensional spacetime is defined as
\begin{eqnarray}\label{Weyltensor0}
	C_{\mu\nu\rho\sigma}=R_{\mu\nu\rho\sigma}-\frac{2}{D-2}( g_{\mu[\rho}R_{\sigma]\nu}-g_{\nu[\rho}R_{\sigma]\mu} )+ \frac{2}{(D-1)(D-2)}R \ g_{\mu[\rho}g_{\sigma]\nu}.
\end{eqnarray}
Here we list some useful formulas.
\begin{eqnarray}\label{Weyltensor}
	C_{z\bar{z}z\bar{z}}&=&e^{4A}C_{1\ z\bar{z}z\bar{z}}+e^{2A}C_{0\ z\bar{z}z\bar{z}},\nonumber\\
	C_{0\ z\bar{z}z\bar{z}}&=&-3T_0+\frac{1}{D-2}\big{[} K_{zmn}K_{\bar{z}}^{mn}-Q_{0\ z\bar{z}}{}^m_{\ m}+6T_0 \big{]}\nonumber\\
	&&-\frac{1}{4(D-1)(D-2)}(3K_{cmn}K^{cmn}-K_cK^c-2Q_{0 c\ m}^{\ \ c\ \ m}+24T_0 )\\
	C_{zi\bar{z}j}&=&e^{2A}C_{1\ zi\bar{z}j}+C_{0\ zi\bar{z}j},\nonumber\\
	C_{0\ zi\bar{z}j}&=&K_{zjl}K_{\bar{z}\ i}^{\ l}-Q_{0\ z\bar{z}ij}\nonumber\\
	&&-\frac{1}{D-2}\big{[} K_{cjl}K^{cl}_{\ i}-\frac{1}{2}K^c K_{cij}-\frac{1}{2}Q_0^{\ c}{}_{\ cij}+g_{ij}(K_{zmn}K_{\bar{z}}^{\ mn}-Q_{0\ z\bar{z}m}^{\ \ \ \ \ \ m}+6T_0) \big{]}\nonumber\\
	&&+\frac{1}{2(D-1)(D-2)}g_{ij}(3K_{cmn}K^{cmn}-K_cK^c-2Q_{0 c\ m}^{\ \ c\ \ m}+24T_0 )\label{Weyltensor1}\\
	C_{ikjl}&=&C_{1\ ikjl}+e^{-2A}C_{0\ ikjl},\nonumber\\
	C_{0\ ikjl}&=&K_{ail}K^a_{\ jk}-K_{aij}K^a_{\ kl}\nonumber\\
	&&-\frac{2}{D-2}\big{[} g_{i[j}R_{0\ l]k}-g_{k[j}R_{0\ l]i}\big{]}\nonumber\\
	&&+\frac{2}{(D-1)(D-2)}g_{i[j}g_{l]k}(3K_{cmn}K^{cmn}-K_cK^c-2Q_{0 c\ m}^{\ \ c\ \ m}+24T_0 )\label{Weyltensor2}\\
	R_{0\ ij}&=&2K_{aim}K^{am}_{\ \ \ j}-K^aK_{aij}-Q_{0a}^{a}{}_{ij}
\end{eqnarray}

Let us focus on the case of \cite{Safdi} with $K_{aij}=\frac{1}{2}k_{ij}, Q_{0\ z\bar{z}ij}=\frac{1}{4}q_{ij}$ and $D=6$. We have
\begin{eqnarray}\label{WeyltensorSafdi}
	C_{0\ z\bar{z}z\bar{z}}&=&\frac{1}{80}( 2k_{mn}k^{mn}+k^2-3q )-\frac{9}{5}t_0\nonumber\\
	C_{0\ zi\bar{z}j}&=&-\frac{1}{8}q_{ij}+\frac{1}{8}k k_{ij}+\frac{1}{80}g_{ij}(k_{mn}k^{mn}-2k^2+q)-\frac{9}{10}t_0 g_{ij}\nonumber\\
	C_{0\ ikjl}&=&(k_{il}k_{jk}-k_{ij}k_{kl})-\frac{1}{2}( g_{i[j}r_{0\ l]k}-g_{k[j}r_{0\ l]i})+\frac{1}{10}g_{i[j}g_{l]k}(3k_{mn}k^{mn}-k^2-2q +24t_0)\nonumber\\
	r_{0\ ij}&=&2k_{im}k^{m}_{\ \ j}-kk_{ij}-q_{ij}
\end{eqnarray}

\end{document}